\documentclass[11pt]{iopart}

\pdfoutput=1 %force axive to use pdflatex

\usepackage{graphicx}
\usepackage{hyperref} %hyperrefs in the bibliography
\usepackage{bbold}
\usepackage{verbatim} %for comment function
\usepackage{cite} %enable multiple citations as e.g. [1-5]
\usepackage{lipsum}

\newcommand{\kb}{k_\mathrm{B}}
\newcommand{\difft}{\frac{\rmd}{\rmd t}}
\newcommand{\opind}[2]{#1_\mathrm{#2}}
\newcommand{\upperind}[2]{{#1}^\mathrm{(#2)}}
\newcommand{\BSig}{\mathbf{\Sigma}}
\newcommand{\llangle}{\langle\hspace{-2.4pt}\langle}
\newcommand{\rrangle}{\rangle\hspace{-2.4pt}\rangle}

\newcommand{\Bllangle}{\Big\langle\hspace{-2.5pt}\Big\langle}
\newcommand{\Brrangle}{\Big\rangle\hspace{-2.5pt}\Big\rangle}

\newcommand{\jhc}{\llangle j\rrangle_\mathrm{hc}}
\newcommand{\jch}{\llangle j\rrangle_\mathrm{ch}}
\newcommand{\rangledis}{\rangle_\mathrm{dis.}}

\newcommand{\trangle}{\rangle_\mathrm{tr}}
\newcommand{\xrangle}{\rangle_\xi}
\newcommand{\ctrangle}{\rangle_{\mathrm{tr},c}}
\newcommand{\cxrangle}{\rangle_{\xi,c}}

\renewcommand{\Re}{\mathop{\mathrm{Re}}}

\renewcommand{\d}{\mathrm{d}}
\newcommand{\ii}{\mathrm{i}}

\begin{document}

%\begin{frontmatter}

\title[Rectification of heat currents across nonlinear quantum chains]{Rectification of heat currents across nonlinear quantum chains: A versatile approach beyond weak thermal contact}
%\date{\today}

\author{T. Motz, M. Wiedmann, J. T. Stockburger and J. Ankerhold}
\address{Institute  for Complex Quantum Systems and IQST, Ulm University - Albert-Einstein-Allee 11, D-89069  Ulm, Germany}
\ead{\mailto{thomas.motz@uni-ulm.de}, \mailto{michael.wiedmann@uni-ulm.de}, \mailto{juergen.stockburger@uni-ulm.de}, \mailto{joachim.ankerhold@uni-ulm.de}}

\begin{abstract}
Within the emerging field of quantum thermodynamics the issues of heat transfer and heat rectification are basic ingredients for the understanding and design of heat engines or refrigerators at nanoscales. Here, a consistent and versatile approach for mesoscopic devices operating with continuous degrees of freedom is developed valid from low up to strong system-reservoir couplings and over the whole temperature range. It allows to cover weak to moderate nonlinearities and is applicable to various scenarios including the presence of disorder and external time-dependent fields. As a particular application coherent one-dimensional chains of anharmonic oscillators terminated by thermal reservoirs are analyzed with particular focus on rectification. The efficiency of the method opens a door  to treat also rather long chains and extensions to higher dimensions and geometries. 
\end{abstract}


\maketitle

\section{Introduction}

Thermodynamics emerged as a theory to describe the properties of systems consisting of macroscopically many degrees of freedom. While it served from the very beginning as a practical tool to quantify the performance of heat engines, it also initiated substantial efforts in fundamental research to pave the road for fields like statistical and non-equilibrium physics. In recent years, triggered by the on-going progress in miniaturizing devices down to the nanoscale, the crucial question if, and if yes, to what extent macroscopic thermodynamics is influenced by quantum mechanics has led to a flurry of literature and the appearance of the new field of quantum thermodynamics.

Of particular relevance as a pre-requisite for the design of actual devices is the understanding of heat transfer and heat rectification.
In macroscopic structures heat transport is typically characterized by normal diffusion such that the heat conductivity is independent of the size of the probe leading to the conventional picture of \emph{local} thermal equilibrium and Fourier's law. The requirements for microscopic models that support this type of normal heat flow has been subject to controversial debates \cite{Lepri1998, Lepri2003, Bonetto2004, Berman2005, Liu2014, Li2015, Dhar2008, Li_Prosen2004} with dimensionality, disorder, and nonlinearities as potential ingredients. 

In the context of mesoscopic physics, it is even less clear under which circumstances the conventional scenario applies. The quantum state of the transport medium may be non-thermal, while energies extracted from or added to thermal baths can still be identified as heat. An extreme case of this type is purely ballistic transport between reservoirs \cite{Nazarov}. 
Thermal rectification occurs when the absolute value of the heat flux through a two-terminal device changes after the temperature difference between the terminals is reversed. Nonlinearities in combination with spatial symmetry breaking are pivotal conditions for the occurrence of rectification for a microscopic modeling \cite{Segal2006,Wu2009}. In a macroscopic, phenomenological setting, a thermal conductivity which depends on space and temperature has been found to be crucial \cite{Peyrard2006}.
 Rectification can be used in thermal diodes or heat valves which have been proposed in classical \cite{Terraneo2002, Li2004, Hu2006, Casati2007, Liu2010, Pereira2013,  Bagchi2017, Romero2017, Kaushik2018} as well as in quantum systems \cite{Segal2005, Segal2005_2, Segal2006, Wu2009, Ruokola2009, Zhang2009, Sanchez2015}. As an alternative to nonlinearities, a coupling of the system to self-consistent reservoirs which guide the constituents of a chain to local equilibrium have been studied \cite{Pereira2011, Pereira2010_2, Bandyopadhyay2011, Pereira2010, Romero_Bastida2014, Pereira2015, Mendonca2015, Guimaraes2015}. Experimentally, thermal rectification has been realized in solid state systems following the ideas of Peyrard \cite{Kobayashi2009, Sawaki2011, Kobayashi2012}. Other implementations for rectification include carbon nanotubes \cite{Chang2006}, trapped ions in optical lattices \cite{Pruttivarasin2011}, hybrid systems where normal metals are tunnel-coupled to superconductors \cite{Martinez-Perez2015} and, recently, superconducting quantum bits \cite{Ronzani2018}. Particularly superconducting circuits allow for a well-controlled modulation of  nonlinearities \cite{Ronzani2018, Shalibo2012, Shalibo2013} and may thus serve as promising candidates for the realization of heat engines operating in the deep quantum regime. In fact, situations where heat transfer is beneficial for the performance of devices have been discussed  for the fast initialization of quantum bits by cooling \cite{Tuorila2017} as well as for properties of heat valves, thermal memories, switches, and transistors \cite{Sanchez2015, Li2012, Vannucci2015, Joulain2016}. 

From the theory point of view, the description of quantum heat transfer is a challenging issue. As part of the broad field of open quantum systems, it became clear in the last years that the underlying processes are extremely sensitive to a consistent treatment of the coupling between system and thermal reservoirs. Lindblad master equations based on \emph{local} dissipators may even violate the second law.
\cite{Levy2014, Stockburger2017}. The `orthodox' approach to Lindblad dissipators suffers from considerable complexity for systems lacking symmetry since it involves all allowed transitions between energy eigenstates. Its perturbative nature typically limits it to systems with weak thermal contact. On the other hand, current experimental activities call for a systematic theoretical approach to quantum heat transfer valid from weak up to strong system-reservoir couplings and down to very low temperatures. The goal of this work is to present such a formulation that, in addition, stands out for its numerical efficiency.  While we here focus on the experimentally relevant case of one-dimensional chains of anharmonic oscillators, generalizations to higher dimensions, more complex geometries, external driving, or set-ups such as heat valves and heat engines are within the scope of the method. The one-dimensional character of the system we study is most clearly maintained by attaching reservoirs at the chain ends only (no scattering to or from transverse channels). 

%, which demands for a careful treatment of system-reservoir interactions and couplings within multipartite systems \cite{Kim2007, Levy2014, Stockburger2017, Pereira2018}. Moreover, due to current experimental activities a systematic approach for quantum heat transfer valid from the weak up to the strong coupling regime and down to very low temperatures is urgently required. The goal of this work is to present such a formulation that, in addition, stands out for its efficiency. While here for specific applications we focus on the experimentally relevant case of one-dimensional chains of anharmonic oscillators, it thus allows for generalizations to higher dimensions, more complex geometries, external driving, or set-ups such as heat valves and heat engines. 

This approach originates from the  description of non-equilibrium quantum dynamics through a stochastic Liouville-von Neumann equation (SLN) \cite{Stockburger2002}, which is an \emph{exact} representation of the formally exact Feynman-Vernon path integral formulation for dissipative quantum systems in terms of a stochastic equation of motion \cite{Stockburger1998, Stockburger2002, Stockburger2003, FeynmanVernon1963}. For reservoirs with ohmic spectral densities and large bandwidths, the SLN can further be simplified to a mixed type of dynamics governed by a stochastic Liouville-von Neumann equation with dissipation (SLED) \cite{Stockburger1999, Gardiner1988}. It has then proven to be particularly powerful to describe systems with continuous degrees of freedom and in presence of external time-dependent driving \cite{Schmidt2011, Schmidt2013}. However, the main challenge to practically implement the SLED is the degrading signal to noise ratio for increasing simulation time. Here, we address this problem by formulating the quantum dynamics in terms of hierarchies of cumulants which are truncated properly. An obvious additional benefit is a vastly improved scaling with system size. 
As we explicitly demonstrate, this treatment provides a very versatile tool to analyze quantum heat transfer in steady state across single or chains of anharmonic oscillators with weak to moderate anharmonicities. A comparison with benchmark data from a direct sampling of the full SLED proves the accuracy of the approach for a broad range of values for the anharmonicity parameter.  
It thus allows to cover within one formulation domains in parameters that are not accessible by alternative approaches  
 \cite{Segal2005, Segal2005_2}. Previous work representing quantum states through cumulants seems mostly limited to closed systems
 \cite{Prezhdo2000, Prezhdo2002, Pereverzev2008, Shigeta2006, Pereverzev2008_2}. However, there is a conceptually related approach including reservoirs \cite{Ruokola2009}, with a focus on smaller structures though. In a classical context, cumulant expansions of fairly high order have been used  \cite{Bricmont2007, Bricmont2007_2}.

The paper is organized as follows: In Section~\ref{sec:stoch_dyn}, we introduce the SLN which represents the basis for the cumulant truncation scheme presented in Section~\ref{sec:cum_trunc}. First applications and comparison with benchmark results are discussed in Section~\ref{sec:state_osci_chain}. The main findings for heat rectification are then presented in Sections~\ref{sec:rect_1HO} and \ref{sec:rect_chains} including an analysis of the physical mechanism that determine the occurrence of rectification. The impact of disorder is considered in Section~ \ref{sec:rect_disordered} before a summary and outlook is given in \ref{sec:sum_out}.

\section{Non-perturbative reduced  Dynamics}
\label{sec:stoch_dyn}

The description of heat transfer is a delicate and challenging issue. 
%It is known that naive applications of standard weak coupling approaches for the reduced dynamics of an aggregate terminated by two thermal reservoirs such as master equations suffer from fundamental inconsistencies \cite{Levy2014,Stockburger2016}. In addition they are restricted to weak aggregate reservoir couplings and elevated temperatures. 
In the sequel we develop an approach which is based on a formally exact formulation of open quantum dynamics derived within a system + bath model. The total Hamiltonian consists of a system Hamiltonian $H_s$, a reservoir Hamiltonian $H_R$ (for the moment we consider a single reservoir), and a system reservoir coupling  $\opind{H}{I} = -qX$, where the latter captures a bilinear coupling of a system's coordinate $q$ and a collective bath coordinate $X$. Due to the macroscopically many degrees of freedom of the reservoir, the fluctuations of the latter in thermal equilibrium can assumed to be Gaussian.\newline

Then, as shown previously, the formally exact path integral representation of the dynamics of the reduced density $\rho(t)={\rm Tr}_{\rm R}\{W(t)\}$ with $W(t)$ being the time evolved density operator in full Hilbert space, has an equivalent representation in terms of a stochastic Liouville-von Neumann equation (SLED) \cite{Stockburger1999}, i.e., 
\begin{equation}
\difft\rho_\xi = \mathcal{L} \rho_\xi =-\frac{\ii}{\hbar}[H_s,\rho_\xi] + \frac{\ii}{\hbar}\xi(t)[q,\rho_\xi]-\frac{\ii}{\hbar}\frac{\gamma}{2}[q,\{p,\rho_\xi\}]\nonumber\;.
\label{eq:SLED}
\end{equation}
Here,  $\xi(t)$ denotes a c-valued noisy driving force whose auto-correlation reconstructs the real part of the bath quantum correlation $L(t-t')$%
\begin{equation}
\langle\xi (t)\xi (t')\rangle = \Re L(t-t')\;,
\label{eq:noisecor}
\end{equation}
where
\begin{equation}
L(t) = \frac{\hbar}{\pi}\int_0^\infty\d \omega J(\omega)[\coth\Big(\frac{\beta\hbar\omega}{2}\Big)\cos(\omega t) - \ii\sin(\omega t)]\,
\label{eq:bath_corr}
\end{equation}
with inverse temperature $\beta=1/(\kb T) $ and the spectral density $J(\omega)$. The latter we assume to be ohmic with a large cut-off frequency $\omega_c$ and coupling constant $\gamma$ acting on a system with mass $m$, i.e.,  
\begin{equation}
J(\omega)=m\frac{\gamma\omega}{[1+(\omega/\omega_c)^2)]^2}\, .
\label{eq:spectralden}
\end{equation}
In this regime (large cut-off), the imaginary part of the correlation function Im$L(t)$ collapses to a $\delta$-function and is accounted for by the $\gamma$-dependent in (\ref{eq:SLED}). We note in passing that Gardiner has identified this equation as the adjoint of a quantum Langevin equation \cite{Gardiner1988, Ford1965}. The random density $\rho_\xi$ propagated according to (\ref{eq:SLED}) by itself lacks of a physical interpretation, while only the expectation value of $\rho=\langle \rho_\xi\xrangle$ represents the physical reduced density of the system.

The SLED is particularly suited to capture the dynamics of open quantum systems with continuous degrees of freedom and has been explicitly applied so previously in various contexts \cite{Stockburger2017, Wiedmann2016}. In the sequel, we follow a somewhat different route though by exploiting
that the multiplicative noise $\xi$ in the SLED turns into additive noise if an adjoint equation governed by $\mathcal{L}^\dagger$ \cite{Breuer} for the dynamics of Heisenberg operators $A$ is used, i.e.,   
\begin{eqnarray}
\difft A_\xi = & \frac{\rmi}{\hbar}[H_s,A_\xi] - \frac{\rmi}{\hbar}\xi(t)[q,A_\xi] + \frac{\rmi}{\hbar}\frac{\gamma}{2}\{p,[q,A_\xi]\}\,.\phantom{.....}
\label{eq:adjoint_sled}
\end{eqnarray}
Expectation values are then obtained from a quantum mechanical average $\langle \cdot\trangle = \mathrm{tr}(\cdot\rho_\xi)$ and a subsequent noise average $\langle \cdot \rangle_\xi$, i.e., 
\begin{equation}
\llangle A\rrangle = \llangle A\trangle\xrangle\;.
\label{eq:doubleav_mom}
\end{equation}
Now, to avoid an explicit noise sampling with its inherent degradation of the signal to noise ratio for longer simulation times, we do not explicitly work with (\ref{eq:adjoint_sled}) but rather derive from it sets of equations of motion for expectation values \cite{Schmidt2011, Schmidt2013}. This way, one arrives at a very efficient scheme to construct the open system dynamics for ensembles of anharmonic oscillators also in regimes which are not accessible by perturbative methods for open quantum systems.

A central element of this formulation is the covariance of two operators $A$ and $B$ which can be transformed to \cite{Stockburger2017, Motz2017}
\begin{eqnarray}
\mathrm{Cov}(A,B) &  = & \textstyle{\frac{1}{2}}\llangle AB+BA\rrangle - \llangle A\rrangle\llangle B\rrangle\nonumber\\
& = & \textstyle{\frac{1}{2}}\llangle AB+ BA\rrangle - \llangle A\opind{\rangle}{tr}\langle B\opind{\rangle}{tr}\xrangle
\nonumber\\
& + &\llangle A\opind{\rangle}{tr}\langle B\opind{\rangle}{tr}\xrangle - \llangle A\rrangle\llangle B\rrangle\\
& = &\langle\mathrm{Cov}_\mathrm{tr}(A,B)\xrangle + \mathrm{Cov}_\xi(\langle A\opind{\rangle}{tr},\langle B\opind{\rangle}{tr})\,
\label{eq:covsplit}
\end{eqnarray}
and thus provides a separation into two covariances, one with respect to the trace average and one with respect to the noise average. Hence, choosing $A$ and $B$ as elements of the operator valued vector $\vec{\sigma}=(q,p)^t$ carrying position and momentum operators, the corresponding covariance matrix $\BSig$  takes the form
\begin{eqnarray}
\BSig =& \upperind{\BSig}{mtr} + \upperind{\BSig}{msc}\;,\\
\label{eq:Sigm_decomp}
\upperind{\BSig}{mtr}_{jk} =&
\langle\mathrm{Cov}_\mathrm{tr}(\sigma_j,\sigma_k)\xrangle\;,\\
\upperind{\BSig}{msc}_{jk} =&
\mathrm{Cov}_\xi(\langle \sigma_j\opind{\rangle}{tr},\langle \sigma_k\opind{\rangle}{tr})\,.
\label{eq:Sigm_parts}
\end{eqnarray}
Technically, this decomposition requires to carefully distinguish between \textit{trace moments} and \textit{trace cumulants}. Namely, the elements of $\BSig$ contain noise expectation values of \textit{trace moments}, while $\upperind{\BSig}{mtr}$ contain noise expectation values of  \textit{tracecovariances}. For general operators $A, B$ we thus introduce  the compact notation for noise expectation values of tracecovariances $\langle\mathrm{Cov}_\mathrm{tr}(A,B)\xrangle$:
\begin{equation}
\llangle AB\rrangle_c = \llangle AB\ctrangle\xrangle\,,
\label{eq:doubleav_cum}
\end{equation}
with $\langle AB\ctrangle=\langle AB\trangle -\langle A\trangle\langle B\trangle$. We also emphasize the equality of the \emph{first} moments and cumulants
\begin{equation}
\langle A\ctrangle = \langle A\trangle\, \ \ , \ \  \llangle AB\rrangle_c=\llangle AB\ctrangle\cxrangle\, .
\label{eq:doubleav}
\end{equation}
With these tools at hand, we can now proceed to develop the approach for nonlinear oscillators in detail.

\section{Cumulant formulation of open quantum dynamics for nonlinear oscillators}
\label{sec:cum_trunc}

\begin{figure}

\begin{center}
\includegraphics[width=7.5cm]{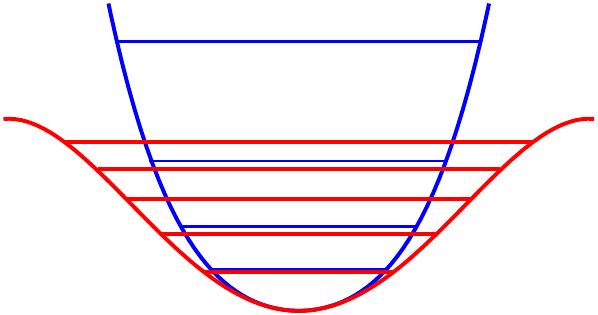}
\end{center}
\caption{Schematic illustration of the considered anharmonic potential $V(q) = \frac{1}{2}m\omega^2q^2+\frac{1}{4}m\kappa q^4$ with the level spacings for $\kappa>0$ (blue) and $\kappa<0$ (red) which are not equidistant. For $\kappa<0$, only metastable states in the vicinity of $q=0$ are considered. The tunneling rates out of the region of the local minimum are supposed to be small and only sufficiently low temperatures which do not induce fluctuations beyond the barrier are considered.}
\label{fig:pot_anharm}
\end{figure}

We will start to consider a single anharmonic oscillator to arrive at a formulation that can then easily be generalized to chains of oscillators.
As a paradigmatic model for a nonlinear oscillator of mass $m$ we choose
\begin{equation}
H_s = \frac{p^2}{2m}+\frac{1}{2}m\omega^2q^2+\frac{1}{4}m\kappa q^4\,,
\label{eq:Hamonean}
\end{equation}
with fundamental frequency $\omega$ and anharmonicity parameter  $\kappa$, see figure.~\ref{fig:pot_anharm}, which can be both positive (stiffer mode) or negative (softer mode). Our main interest is in weakly anharmonic systems, i.e., even for
negative $\kappa$, where the potential is not bounded from below, resonances are long-lived and may be treated approximately as eigenstates of the Hamiltonian. This is valid if thermalization through external reservoirs is fast compared to the resonance lifetime.

While for a purely linear system ($\kappa=0$) a formulation in terms of cumulants leads to closed equations of motion for the first and second order cumulants \cite{Schmidt2011, Schmidt2013}, this is no longer the case for nonlinear oscillators. The challenge is thus to implement a systematic procedure for the truncation of higher order cumulants which on the one hand provides an accurate description for weak to moderate anharmonicity parameters and on the other hand leads to an efficient numerical scheme also for ensembles of those oscillators. We emphasize that we are primarily interested in quantum heat transfer in steady state situations and not in the full wealth of nonlinear dissipative quantum dynamics. Accordingly, in a nutshell, a formulation which satisfies both criteria factorizes all higher than second order moments (Wick theorem) and allows to capture anharmonicities effectively by state dependent frequencies which must be determined self-consistently.  This perturbative treatment of nonlinearities for quantum oscillators has some similarities in common with the one for classical systems \cite{Landau} but turns out to be much more involved due to the highly complex equation of motion (\ref{eq:adjoint_sled}) as we will show now. 

For this purpose, for systems, where the relevant dynamics occurs around a single potential minimum, it is convenient to set initial values of operators $A$  equal to zero and use
\begin{equation}
\llangle A\rrangle = 0
\label{eq:doubleav_zero}
\end{equation}
henceforth. Then, starting from (\ref{eq:adjoint_sled}) one obtains two coupled equations for position and momentum according to
\begin{eqnarray}
\frac{\rmd}{\rmd t}\langle q\ctrangle &=\frac{1}{m}\langle p\ctrangle\nonumber\\
\frac{\rmd}{\rmd t}\langle p\ctrangle &=-m\omega^2\langle q\ctrangle-m\kappa\langle q^3\trangle -\gamma\langle p\ctrangle +\xi(t)\,.
\label{eq:first_moments_ex}
\end{eqnarray}
Here,  the third trace moment can be expressed as a linear combination of products of cumulants
\begin{equation}
\langle q^3\trangle = \langle q^3\ctrangle + 3\langle q^2\ctrangle\langle q\ctrangle +\langle q\ctrangle^3\,.
\label{eq:thirdmom_cum}
\end{equation}
This kind of transformation represents a systematic separation and summation over all possible subsets of trace averaged products, for details see~\cite{Kubo1962}.\newline

\textit{a) Truncation of trace and noise cumulants}\newline

A straightforward approach to treat higher than second order moments is to assume the approximate validity of Wick's theorem and neglect all higher than second order cumulants so that (\ref{eq:thirdmom_cum}) reduces to \begin{equation}
\langle q^3\trangle \approx 3\langle q^2\ctrangle\langle q\ctrangle +\langle q\ctrangle^3\, .
\label{eq:thirdmom_cumtrunc}
\end{equation}
This procedure does not immediately lead to a closure of (\ref{eq:first_moments_ex}) although with parametric driving through  $\langle q^2\ctrangle$, which couples to the equations for the second cumulants which we analyze later.

As shown for the linear system, the covariance matrix consists of two parts, where the elements of one part is given by the noise averaged product $\llangle A\ctrangle\langle B\ctrangle\xrangle$ \cite{Motz2017}. To handle these terms, we derive the equations of motion for these products with $\mathcal{L}^\dagger$ and turn the noise moments to \textit{noise cumulants} which is trivial for all linear contributions since $\llangle A\ctrangle\cxrangle=\llangle A\rrangle=0$ and also for the system- bath correlation which contains the quantum noise whose expectation value is zero. The resulting equations of motion then read
\begin{eqnarray}
\frac{\rmd}{\rmd t}\llangle q\ctrangle\langle q\ctrangle\cxrangle =&\frac{2}{m}\llangle q\ctrangle\langle p\ctrangle\cxrangle\nonumber\\
\frac{\rmd}{\rmd t}\llangle p\ctrangle\langle p\ctrangle\cxrangle =&-2m\omega^2\llangle q\ctrangle\langle p\ctrangle\cxrangle\nonumber\\
&-2m\kappa[3\llangle q^2\ctrangle\langle q\ctrangle\langle p\ctrangle\rangle_{\xi} +\llangle q\ctrangle^3\langle p\ctrangle\xrangle]\nonumber\\ 
&-2\gamma\llangle p\ctrangle\langle p\ctrangle\cxrangle +2\langle\xi(t)\langle p\trangle\cxrangle\nonumber\\
\frac{\rmd}{\rmd t}\llangle q\ctrangle\langle p\ctrangle\cxrangle =&\frac{1}{m}\llangle p\ctrangle\langle p\ctrangle\cxrangle-m\omega^2\llangle q\ctrangle\langle q\ctrangle\cxrangle\nonumber\\
&-m\kappa[3\llangle q\ctrangle^2\langle q^2\ctrangle\xrangle+\llangle q\ctrangle^4\xrangle]\nonumber\\
&-\gamma\llangle q\ctrangle\langle p\ctrangle\cxrangle +\langle\xi(t)\langle q\trangle\cxrangle\,.
\label{eq:sigma_stoch_tr}
\end{eqnarray}
In equivalence to the equations of the trace cumulants, here the contributions which can not be turned immediately from moments to cumulants are the ones which enter from the nonlinearity of the system. The second and third equations in (\ref{eq:sigma_stoch_tr}) contain higher order moments with respect to the noise average. The trace cumulants in these moments constitute c-numbers, and hence, the order of the higher order noise moments is determined by the sum of the exponents outside of the trace averages $\langle .\ctrangle$.  Hence,  $\llangle q^2\ctrangle\langle A\ctrangle\langle B\ctrangle\rangle_{\xi}$ ($A,B$=$q,p$) constitutes a third order moment with respect to the noise and  $\llangle q\ctrangle^3\langle A\ctrangle\xrangle$ a fourth order moment. These moments as linear combinations of cumulants are
\begin{equation}
\llangle q^2\ctrangle\langle A\ctrangle\langle B\ctrangle\rangle_{\xi}=\, \llangle q^2\ctrangle\langle A\ctrangle\langle B\ctrangle\cxrangle+\llangle A\ctrangle\langle B\ctrangle\cxrangle\llangle q^2\rrangle_c\nonumber\,\\
\label{eq:full_noise_cums_1}
\end{equation}
and 
\begin{equation}
\llangle q\ctrangle^3\langle A\ctrangle\xrangle =\, \llangle q\ctrangle^3\langle A\ctrangle\cxrangle +3\llangle q\ctrangle\langle A\ctrangle\cxrangle\llangle q\ctrangle\langle q\ctrangle\cxrangle\,,
\nonumber\\
\label{eq:full_noise_cums_2}
\end{equation}
where $\llangle A\rrangle_c=0$ is already considered.

Setting the third- and fourth-order cumulant to zero gives 
\begin{equation}
\llangle q^2\ctrangle\langle A\ctrangle\langle B\ctrangle\rangle_{\xi}\approx\,\llangle A\ctrangle\langle B\ctrangle\cxrangle\llangle q^2\rrangle_c=\llangle A\trangle\langle B\trangle\xrangle\llangle q^2\rrangle_c
\label{eq:approx_noise_cums1}
\end{equation}
and
\begin{eqnarray}
\llangle q\ctrangle^3\langle A\ctrangle\xrangle &\approx 3\llangle q\ctrangle\langle A\ctrangle\cxrangle\llangle q\ctrangle\langle q\ctrangle\cxrangle\nonumber\\
&=3\llangle q\trangle\langle A\trangle\xrangle\llangle q\trangle\langle q\trangle\xrangle\,,
\label{eq:approx_noise_cums2}
\end{eqnarray}
where the equalities in (\ref{eq:approx_noise_cums1}) and (\ref{eq:approx_noise_cums2}) follow directly from (\ref{eq:doubleav}). With (\ref{eq:approx_noise_cums1}), products of the noise expectation values of the \textit{first} and \textit{second} trace cumulants enter. The dynamics of the second trace cumulants is shown in \ref{app:2ndtrace_cumulants}, where an analysis of the steady-state shows that the fluctuations induced by the coupling to the first trace moments are exponentially suppressed and, therefore, this second trace cumulants have a stable fixed point at zero. This leads to a decoupling of the equations for the first- and second trace cumulants and a reduction of the covariance matrix to $\BSig = \upperind{\BSig}{msc}$. This allows us to simplify the notation by using $\llangle A\trangle\langle B\trangle\xrangle=\llangle AB\rrangle$ which is valid for steady-states. The elements of $\BSig$ are then determined by a system of differential equations
\begin{eqnarray}
\difft \llangle q^2\rrangle =&\frac{2}{m}\llangle qp\rrangle\nonumber\\\difft \llangle p^2\rrangle =&-2m\tilde{\omega}^2\llangle qp\rrangle-2\gamma\llangle p^2\rrangle+2\langle\xi(t)\langle p\trangle\xrangle\nonumber\\
\difft\llangle qp\rrangle =&\frac{1}{m}\llangle p^2\rrangle-m\tilde{\omega}^2\llangle q^2\rrangle-\gamma\llangle qp\rrangle+\langle\xi(t)\langle q\trangle\xrangle\,,
\label{eq:stoch_mat}
\end{eqnarray}
which represent the steady-state by algebraic equations if the left hand sides are set to zero and the system-bath correlations $\langle\xi(t)\langle \sigma_j\trangle\xrangle$ are in their respective steady-state value. We introduced the effective frequency
\begin{equation}
\tilde{\omega}^2=\omega^2+3\kappa\llangle q^2\rrangle\,,
\label{eq:eff_freq}
\end{equation} 
Since $\llangle q^2\rrangle$ is an element of the covariance but enters also the effective frequency, our approach can be seen as a type of self-consistent mean-field formulation with some similarities to the one developed by Ruokola et al. in \cite{Ruokola2009}. We note in passing that  state-dependent frequencies are also known from classical perturbative treatments of anharmonic oscillators \cite{Landau}. In the quantum model considered here, the expectation value of the amplitude is zero ($\llangle q\rrangle =0$) but the width of the state $\llangle q^2\rrangle$ enters into the effective frequency. The noise average leads to a temperature dependency of $\llangle q^2\rrangle$ and therefore to an effective frequency $\tilde{\omega}$ which depends on the temperature of the heat bath. In case of oscillators interacting with two thermal reservoirs at different temperatures, the situation we consider below, the effective frequency does not only depend on both of these temperatures but also on cross-correlations between system and reservoirs.

A direct calculation of the dynamics of the full covariance matrix $\BSig$ reveals a dependence on higher order moments which are given by $\llangle q^4\rrangle$ and $\Bllangle\frac{q^3p+pq^3}{2}\Brrangle$. The presented formalism represents a transformation of the moments contained in the covariance matrix into cumulants and an expansion with subsequent truncation of higher order cumulants. This procedure is equivalent to an approximation of the higher order moments given by
\begin{eqnarray}
&\llangle q^4\rrangle \approx 3\llangle q^2\rrangle^2\nonumber\\
&\Bllangle\frac{q^3p+pq^3}{2}\Brrangle \approx 0\,.
\label{eq:moment_approx}
\end{eqnarray}
The first equation is obviously the content of Wick's theorem, just like the second equation if the system is in steady-state where $\Bllangle\frac{qp+pq}{2}\Brrangle=0$ holds. Nevertheless, the truncation scheme we present here provides a systematic approach that accounts for both, the quantum and the thermal fluctuations. Therefore it represents an extension of previous schemes based on truncations of higher order moments or cumulants for closed systems \cite{Prezhdo2000, Prezhdo2002, Pereverzev2008, Shigeta2006, Pereverzev2008_2}.

\bigskip
\textit{b) System-bath correlation function}\newline

For a compact treatment of the system-bath correlations and a subsequent extension to multipartite systems we treat the system as a linear one with effective frequency and use a matrix notation of the derived steady-state equations (\ref{eq:stoch_mat}) via a Lyapunov equation as in previous work \cite{Stockburger2016, Motz2017}:
\begin{equation}
\mathbf{M}\BSig + \BSig\mathbf{M}^\dagger\  + \mathbf{y}^\dagger + \mathbf{y} = 0\,.
\label{eq:sigma_noisesteady}
\end{equation}
The matrices have dimension $2N\times 2N$ with $N$ being the number of oscillators. $\mathbf{M}$ contains details of the model and the damping and reads for one oscillator 
\begin{equation}
\mathbf{M}=
\left(\begin{array}{cc}
0 & 1/m\\
-m\tilde{\omega}^2 & -\gamma\\
\end{array}\right)\,.
\label{eq:M_mat}
\end{equation}
with the effective frequency $\tilde{\omega}^2=\omega^2+3\kappa\llangle q^2\rrangle$ which itself depends on an element of the covariance matrix $\BSig$. Therefore, (\ref{eq:sigma_noisesteady}) is solved self-consistently with the solution for the linear system ($\kappa=0$) as an initial guess for $\mathbf{M}$ and the 
system-bath correlations. These are shifted to the matrix $\mathbf{y}$ which reads
\begin{equation}
\mathbf{y}(t)=
\left(\begin{array}{cc}
0 & \langle\xi(t)\langle q\trangle\xrangle\\
0 & \langle\xi(t)\langle p\trangle\xrangle\\
\end{array}\right)\,.
\label{eq:y_mat}
\end{equation}
In terms of a tensor product of the vector carrying the phase space variables $\vec{\sigma}=(q,p)^t$ and a vector with the noise $\vec{\xi}=(0,\xi)^t$, the correlation matrix is 
\begin{equation}
\mathbf{y}(t) = \llangle\vec{\sigma}\opind{\rangle}{tr}\vec{\xi}^t(t)\xrangle\,.
\label{eq:sysbathcorr}
\end{equation} 
If the effective frequency is supposed to be fixed, the equations (\ref{eq:stoch_mat}) can considered as a linear set of equations with additional driving $\xi(t)$ which are formally solved by the Green's function: 
\begin{equation}
\langle\vec{\sigma}\rangle_\mathrm{tr} = \int_0^t\rmd t'\mathbf{G}(t-t')\vec{\xi}(t')\,.
\label{eq:formsol}
\end{equation} 
The noise expectation value of the product of $\langle\vec{\sigma}\rangle_\mathrm{tr}$ with the noise vector $\vec{\xi}^t(t)$ gives a form of $\mathbf{y}$ where all noises are shifted to the bath auto-correlation function:
\begin{equation}
\mathbf{y}(t) = \mathbf{y}(0) + \int_0^t\rmd t' \mathbf{G}(t-t')\langle\vec{\xi}(t)\vec{\xi}^t(t')\xrangle\;.
\label{eq:yfunc_int}
\end{equation}
The elements of the matrix $\mathbf{L'}(t-t')=\langle\vec{\xi}(t)\vec{\xi}^t(t')\xrangle$ are given by (\ref{eq:bath_corr}) which completes the presented formalism to a deterministic description based solely on model parameters. For consistent results with (\ref{eq:sigma_noisesteady}) one has to integrate to sufficiently large times until a constant steady-state value of $\mathbf{y}$ is reached.  By usage of the time-translational symmetry of the system, this integral can be derived with respect to time and written as a system of two coupled differential equations
\begin{eqnarray}
\dot{\mathbf{y}}(t) =& \mathbf{G}(t)\mathbf{L}(t) \label{eq:yfuncm}\\
\dot{\mathbf{G}}(t) =& \mathbf{M}\mathbf{G}(t)\,,
\label{eq:greenfunc}
\end{eqnarray}
which allows a simple and computationally efficient implementation. Treating both, the system-bath correlation and the dissipative parts as a linear system with effective frequency $\tilde{\omega}$ provides a consistent formalism for the nonlinear system. 
\newline

\textit{c) Generalization to chains of oscillators}\newline

\label{sec:chains_ext}
\begin{figure}
\begin{center}
\includegraphics[width=7.5cm]{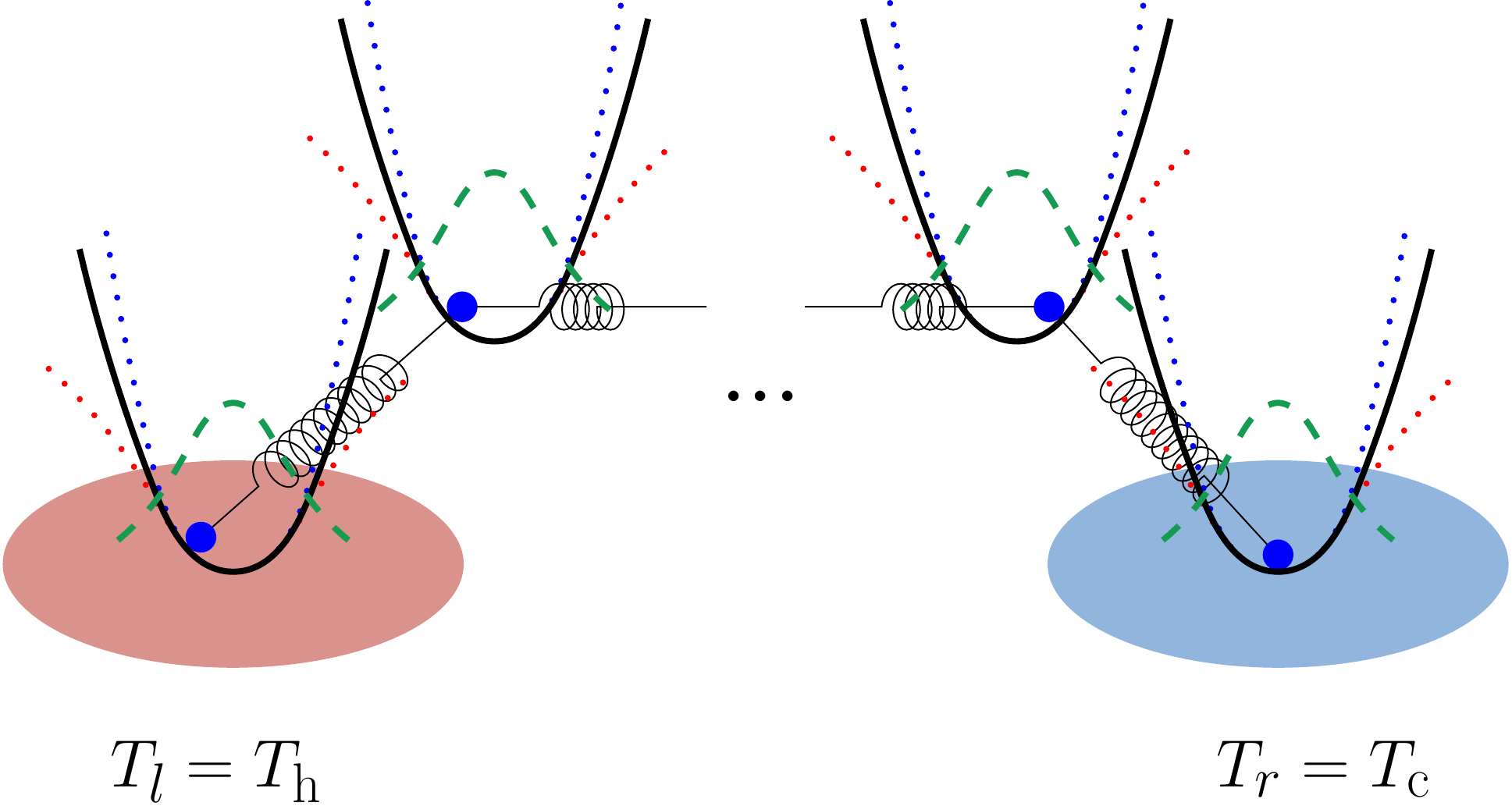}
\end{center}
\caption{Schematic diagram of a chain of anharmonic oscillators which is terminated by thermal reservoirs. The dotted lines illustrate the on-site potentials for $\kappa>0$ (blue) and $\kappa<0$ (red), while the solid black line represents the harmonic case with $\kappa=0$. The width of the potential determines the quantum state (green) with its effective frequency $\tilde{\omega}_n$.}
\label{fig:chain_anh}
\end{figure}

In order to step forward to chains of anharmonic oscillators, we choose a quadratic coupling between adjacent entities, i.e., 
\begin{equation}
H_s = \sum_{n=1}^N\frac{p_n^2}{2m}+\frac{1}{2}m\omega_n^2 q_n^2+\frac{1}{4}m\kappa q_n^4+\frac{\mu}{2}\sum_{n=1}^{N-1}(q_n-q_{n+1})^2\;
\label{eq:Ham_qu}
\end{equation}
and, for the sake of transparency, work with chains with homogeneous mass $m$, anharmonicitiy $\kappa$, and inter-oscillator coupling $\mu$. If the on-site frequencies are also homogeneous, we denote them with $\omega$. The generalization to inhomogeneous structures is straighforward (see Sec.~\ref{sec:rect_disordered}).
Note that this Hamiltonian differs from the paradigmatic Fermi-Pasta-Ulam model \cite{Berman2005} in that there nonlinearities appear in the inter-oscillator couplings while here they feature on-site pinning type of potentials. 

Now, to study quantum heat transfer through this extended structure, the formulation for a single oscillator is easily generalized along the lines presented in \cite{Stockburger2017, Motz2017}. The chain is terminated at its left ($l$) and right ($r$) end by two independent reservoirs, see figure.~\ref{fig:chain_anh}, with bath correlations given by 
\begin{equation}
\langle\xi_\nu(t)\xi_{\nu'}(t')\rangle =\delta_{\nu,\nu'}\Re L_\nu(t-t')\,, \; \nu=l, r
\label{eq:noisecor_multres}
\end{equation}
where the $L_\nu(t)$ are specified in (\ref{eq:bath_corr}) and (\ref{eq:spectralden}) with $J(\omega)\to J_\nu(\omega)$ according to $\gamma\to \gamma_\nu$ and $\beta\to \beta_\nu=1/k_{\rm B}T_\nu$. For heat transfer one considers situations with $T_l\neq T_r$ corresponding to a `cold' and a `hot' reservoir with temperatures $T_h>T_c$. 

In absence of cross-correlations, the contributions of the two reservoirs are additive in the corresponding Caldeira-Leggett Hamiltonian \cite{Caldeira1983a} so that we have in generalization of (\ref{eq:adjoint_sled})
\begin{eqnarray}
\difft A_\xi = & \frac{\rmi}{\hbar}[H_s,A_\xi] - \frac{\rmi}{\hbar}\xi_l(t)[q_1,A_\xi] - \frac{\rmi}{\hbar}\xi_r(t)[q_N,A_\xi]\nonumber\\
& + \frac{\rmi}{\hbar}\frac{\gamma_l}{2}\{p_1,[q_1,A_\xi]\} + \frac{\rmi}{\hbar}\frac{\gamma_r}{2}\{p_N,[q_N,A_\xi]\}\,.
\label{eq:adjoint_sled_chain}
\end{eqnarray}
Accordingly, the cumulant truncation goes through as for a single oscillator and the dynamics of the phase space operators collected in the operator-valued vector $\vec{\sigma}=(q_1,p_1,\dots,q_N,p_N)^t$ follows from
\begin{equation}
\difft\langle\vec{\sigma}\rangle_\mathrm{tr}=\mathbf{M}\langle\vec{\sigma}\rangle_\mathrm{tr}+\vec{\xi}(t)\,.
\label{eq:1st_cumulants}
\end{equation}
Here $\vec{\xi}=(0, \xi_l, 0, \dots, 0, \xi_r)^t$ and the matrix $\mathbf{M}$ contains the dissipative parts and the effective frequencies $\tilde{\omega}_n$ according to (\ref{eq:eff_freq}) to effectively account for the nonlinearities and to be determined self-consistently.
The detailed structure of $\mathbf{M}$ for the Hamiltonian (\ref{eq:Ham_qu}) is shown in \ref{app:system_matrix}. Having $\mathbf{M}$ at hand, the multipartite system obeys the steady-state equations (\ref{eq:sigma_noisesteady}), (\ref{eq:yfuncm}), and (\ref{eq:greenfunc}), where the only non-zero elements in $\mathbf{L}(t-t')$ are the second and last diagonal elements containing the auto-correlation functions $L_l(t)$ and $L_r(t)$, respectively.

\bigskip
\textit{d) Energy fluxes}\newline

Temperature differences between thermal reservoirs at different ends of the chain give rise to an energy flux and thus heat transfer. 
Locally, for the change of energy at an individual site $n$ one must distinguish between the oscillators at the boundaries ($n=1, N$) and the oscillators in the bulk. For the latter, the change in energy follows from
\begin{equation}
H_n = \frac{p_n^2}{2m}+\frac{1}{2}m\omega_n^2 q_n^2+\frac{1}{4}m\kappa q_n^4+\frac{\mu}{2}(q_{n-1}-q_n)^2 + \frac{\mu}{2}(q_n-q_{n+1})^2\,
\end{equation}
and includes also the nearest neighbor coupling, while for the oscillators at the boundaries only one coupling term appears and the other one is replaced by the coupling to the respective reservoir. Calculating the time evolution for each of these parts according to $\difft \langle H_n\trangle = \langle\mathcal{L}^\dagger H_n\trangle$  leads, after performing the stochastic average, to the dynamics of the on-site energies. This then directly leads to the energy fluxes between adjacent sites and between the oscillators at the ends and the respective reservoirs, namely, 
\begin{eqnarray}
\difft \llangle H_1\rrangle &=& \llangle j_{l,1}\rrangle - \llangle j_{1,2}\rrangle \nonumber\\
\difft \llangle H_n\rrangle &=& \llangle j_{n-1,n}\rrangle-\llangle j_{n,n+1}\rrangle,\quad 2\leq n\leq N-1\nonumber\\
\difft \llangle H_N \rrangle &=& \llangle j_{N-1,N}\rrangle - \llangle j_{r,N}\rrangle\,, \nonumber\\
\end{eqnarray}
where respective heat fluxes follow from
\begin{eqnarray}
\llangle j_{n-1,n}\rrangle &=& \frac{\mu}{m}\llangle q_{n-1}p_n\rrangle,\quad\phantom{-} 2\leq n\leq N\label{eq:heatfluxleft}\\
\llangle j_{n,n+1}\rrangle &=&  -\frac{\mu}{m}\llangle q_{n+1}p_n\rrangle,\quad 1\leq n\leq N-1\label{eq:heatfluxright}\\
\llangle j_{l,1}\rrangle &=& \frac{1}{m}\langle\xi_l(t)\langle p_1\opind{\rangle}{tr}\xrangle - \gamma_l\Bllangle\frac{p_1^2}{m}\Brrangle\label{eq:heatfluxres1}\\
\llangle j_{r,N}\rrangle &=& -\frac{1}{m}\langle\xi_r(t)\langle p_N\opind{\rangle}{tr}\xrangle + \gamma_r\Bllangle\frac{p_N^2}{m}\Brrangle\label{eq:heatfluxresN}\,.
\end{eqnarray}
One should note that the above averages are trace and noise moments which means that the respective currents in the bulk in (\ref{eq:heatfluxleft}) and (\ref{eq:heatfluxright}) are determined by elements of the covariance matrix. The first terms in (\ref{eq:heatfluxres1}) and (\ref{eq:heatfluxresN}) represent system-bath correlations and have to be computed as elements of $\mathbf{y}$ (\ref{eq:sysbathcorr}) to obtain the covariance matrix. Therefore, once the covariance matrix is known, the energy currents are implicitly known as well.

\bigskip
\textit{e) Rectification of heat transfer}\newline

An important measure for the asymmetry in heat transfer is the rectification coefficient. It quantifies the net heat current when the temperature difference in the reservoirs is reversed. The configuration with the left reservoir being  hot at temperature $T_l=T_h$ and the right one being cold at temperature $T_r=T_c<T_h$ is  denoted by $h\rightarrow c$ with the corresponding heat current in steady state $\jhc$. The opposite situation, where {\em only} the temperatures are exchanged, i.e.\ $T_l=T_c<T_r=T_h$, is termed $c\leftarrow h$ with the respective heat current $\jch$. Based on this notation the rectification coefficient reads
\begin{equation}
\alpha = \frac{|\jhc|-|\jch|}{|\jhc|+|\jch|}\, .
\label{eq:rectcoef}
\end{equation}
Since heat always flows from hot to cold, $\alpha>0\ (<0)$ means that the heat flow $T_l\to T_r$ ($T_r\to T_l$) exceeds that from $T_r\to T_l$ ($T_l\to T_r$).  The basic ingredients for finite rectification are {\em both} nonlinearity and spatial symmetry breaking \cite{Segal2006, Wu2009}. A purely harmonic system does not show any rectification even in presence of spatial asymmetry which implies that non-equidistant energy level spacings induced by the nonlinearities are a crucial pre-requisite. In the present case of anharmonic oscillators, nonlinearities are  effectively accounted for by effective state dependent frequencies $\tilde{\omega}_n$ that depend on temperature and damping.

\bigskip
\section{Test of the approach}
\label{sec:state_osci_chain}

\begin{figure*}[]
\begin{minipage}{8.5cm}
\includegraphics[width=7.5cm]{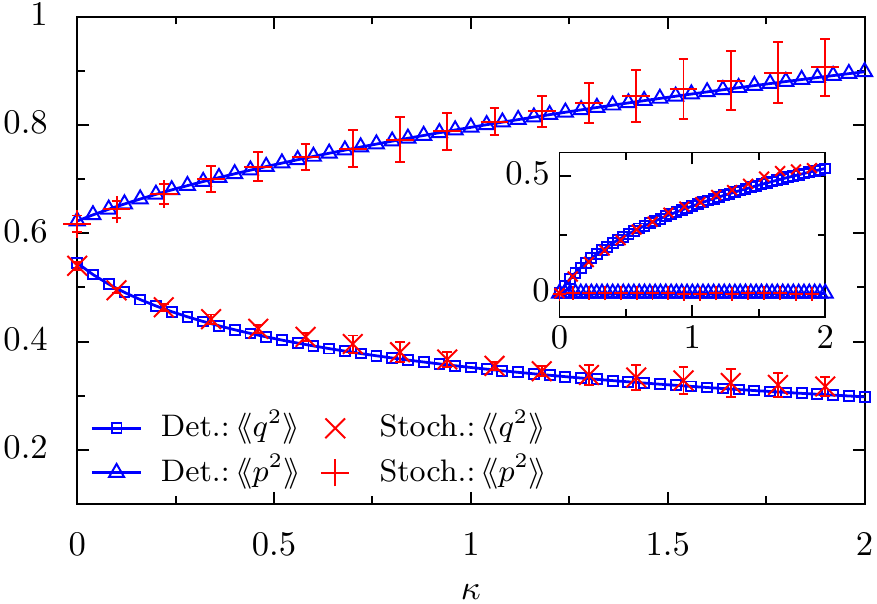}
\end{minipage}
\hfill
\begin{minipage}{8.5cm}
\includegraphics[width=7.5cm]{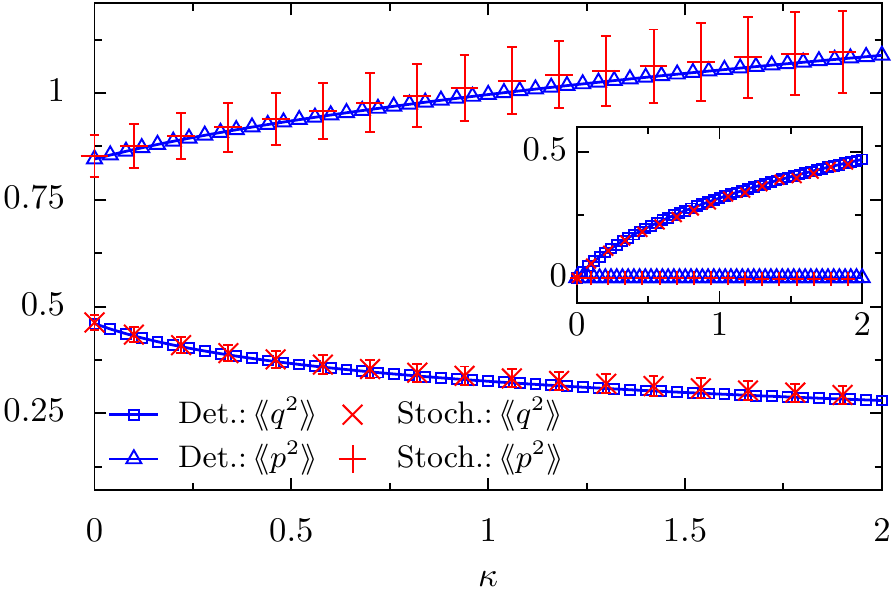}
\end{minipage}
\caption{The diagonal elements of the covariance matrix $\BSig$ versus the anharmonicity $\kappa$ for a single oscillator whose Hamiltonian is given by (\ref{eq:Hamonean}) and which is coupled to a reservoir with $\beta=3$ and damping $\gamma=0.1$ (left) and $\beta=5$ and $\gamma=0.5$ (right). The blue symbols are results from the deterministic cumulant expansion presented here, while the red crosses are obtained from a stochastic sampling of the SLED in position representation. The inset shows the higher order moments (multiplied with the respective $\kappa$) of the exact dynamics obtained by a stochastic sampling of the SLED (red) and the approximations of these moments according to (\ref{eq:moment_approx}) (blue). Other parameters are $\omega=1$, $m=1$, $\hbar=1$ and $\omega_c=30$.}
\label{fig:anham_comp_det_stoch}
\end{figure*}

\textit{a) The classical limit}\newline

As a first and simple illustration of the presented method, we consider an anharmonic oscillator coupled to a single classical thermal reservoir. In thermal equilibrium, (\ref{eq:stoch_mat}) lead to algebraic equations which can be rearranged as
\begin{eqnarray}
\llangle qp\rrangle=& 0\nonumber\\
\llangle p^2\rrangle =& \frac{1}{\gamma}\langle\xi(t)\langle p\trangle\xrangle\nonumber\\
\llangle q^2\rrangle =&\frac{1}{m\tilde{\omega}^2}[\frac{1}{m\gamma}\langle\xi(t)\langle p\trangle\xrangle+\langle\xi(t)\langle q\trangle\xrangle]\,.
\label{eq:app_stoch_mat_final}
\end{eqnarray}
Now, only the system-bath correlations have to be calculated to arrive at a quadratic equation for $\llangle q^2\rrangle$.\newline

In the classical limit with $\omega_c\to\infty$  and $\beta\to 0$, the real part of the bath auto-correlation function (\ref{eq:bath_corr}) reduces to  $L'(t-t')=2\kb Tm\gamma\delta(t-t')$ so that (\ref{eq:yfunc_int}) reads $\mathbf{y}(t) = \kb Tm\gamma \mathbf{G}(0)$;  no initial system-bath correlations are assumed since we take $\mathbf{y}(0)=0$. This way, from $\mathbf{G}(0)=\mathbb{1}$, one arrives at
\begin{equation}
\mathbf{y}(t)=
\left(\begin{array}{cc}
0 & 0\\
0 & y_p\\
\end{array}\right)\,,
\label{eq:y_mat_class}
\end{equation}
with $y_p=\langle\xi(t)\langle p\ctrangle\xrangle=\kb Tm\gamma$ while $\langle\xi(t)\langle q\ctrangle\xrangle=0$. With these findings, (\ref{eq:app_stoch_mat_final}) takes the form known for classical harmonic oscillators \cite{Weiss}, however, with effective frequency $\tilde{\omega}$
\begin{eqnarray}
\llangle qp\rrangle=& 0\nonumber\\
\llangle p^2\rrangle =& \frac{y_p}{\gamma}=m\kb T\nonumber\\
\llangle q^2\rrangle =&\frac{y_p}{(m\tilde{\omega})^2\gamma}=\frac{\kb T}{m\tilde{\omega}^2}\,.
\label{eq:app_stoch_mat_class}
\end{eqnarray}
The quadratic equation for $\llangle q^2\rrangle$ has two solutions with only one being physical, namely, 
\begin{equation}
\llangle q^2\rrangle=\frac{\omega^2}{6\kappa}\Big(1-\sqrt{1+\frac{12\llangle q^2\rrangle_0}{\omega^2}\kappa}\Big)\;.
\label{eq:sol_class}
\end{equation} 
Illuminating is an expansion for small $|\kappa|$ which yields  
\begin{equation}
\llangle q^2\rrangle\approx\llangle q^2\rrangle_0(1-\frac{3\kappa}{\omega^2}\llangle q^2\rrangle_0)\;,
\label{eq:sol_class_taylor}
\end{equation} 
with the harmonic result $\llangle q^2\rrangle_0=\kb T/(m\omega^2)$. As expected, $\kappa>0$ (stiffer mode) decreases $\llangle q^2\rrangle$, while $\kappa<0$ (softer mode) leads to a broadening. (\ref{eq:sol_class}) also provides an upper bound for the validity of the perturbative treatment in case $\kappa<0$,  namely, $|\tilde{\omega}^2-\omega^2|=3 |\kappa| \llangle q^2\rrangle_0/\omega^2 < 1/4$. Indeed, in comparison to an exact numerical calculation of $\llangle q^2\rrangle$, one finds that in this range the result (\ref{eq:sol_class}) provides an accurate description with deviations at most of order 1\%. 

\bigskip
\textit{b) Quantum oscillator in thermal equilibrium}\newline

\begin{figure*}
\begin{minipage}{8.5cm}
\includegraphics[width=7.5cm]{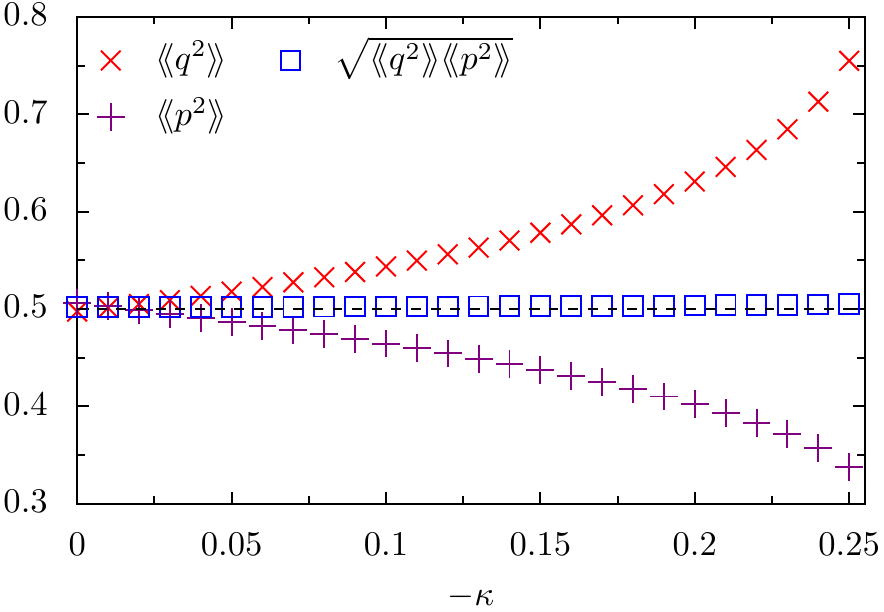}
\end{minipage}
\hfill
\begin{minipage}{8.5cm}
\includegraphics[width=7.5cm]{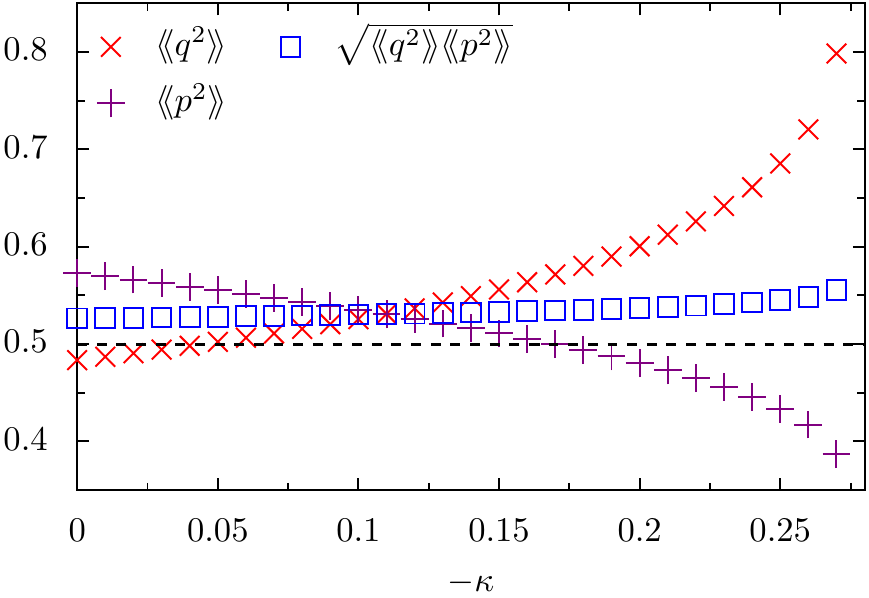}
\end{minipage}
\caption{$\llangle q^2\rrangle$, $\llangle p^2\rrangle$ and the uncertainty $\sqrt{\llangle q^2\rrangle\llangle p^2\rrangle}$ versus the anharmonicity $\kappa$ for a damped anharmonic oscillator with system Hamiltonian (\ref{eq:Hamonean}). Left plot shows $\gamma=0.01$ and the right one $\gamma=0.10$. The dashed lines are a guide for the eye and represent the value $0.5$. Other parameters are $\omega=1$, $m=1$, $\hbar=1$, $\kb T=0$ and $\omega_c=30$.}
\label{fig:anham_neg_kap}
\end{figure*}

\begin{figure*}
\begin{center}
\includegraphics[width=17cm]{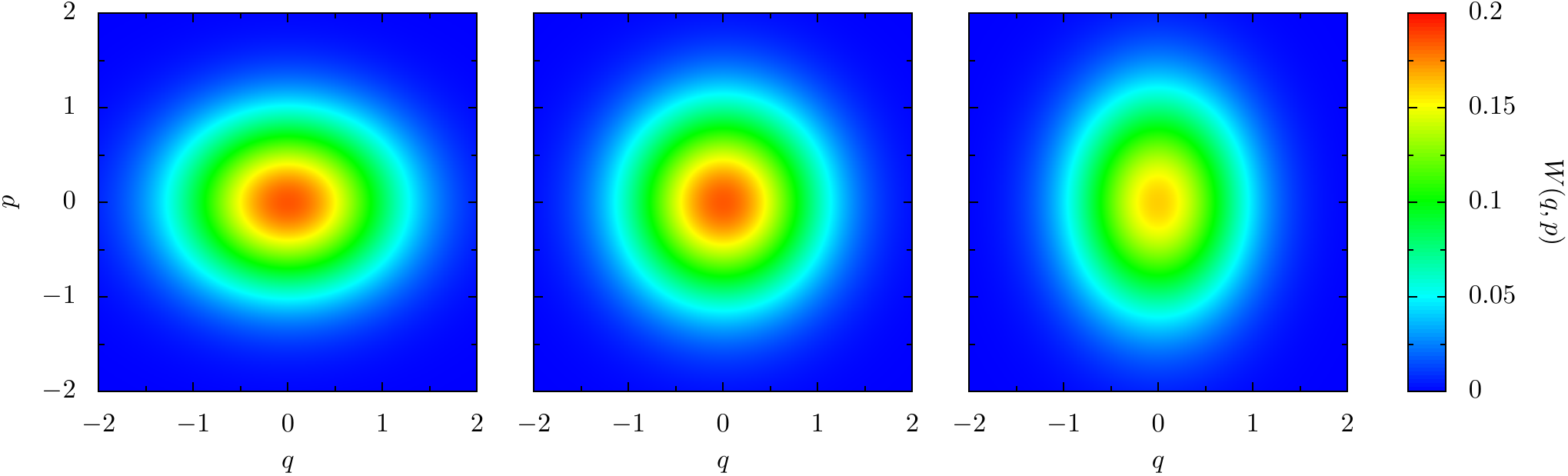}
\end{center}
\caption{Wigner functions of an anharmonic oscillator (\ref{eq:Hamonean}) for different values of $\kappa=-0.2$ (left), $0.0$ (middle) and $0.5$ (right). The plots show the broadening of the state for negative $\kappa$ and the squeezing for positive $\kappa$, while $\kappa=0$ gives an almost circular distribution which is a consequence of small damping $\gamma=0.01$. Other parameters are $\omega=1$, $m=1$, $\hbar=1$, $\kb T=0$ and $\omega_c=30$.}
\label{fig:wigner_T_0}
\end{figure*}

We now proceed to analyze the performance of the approach in the quantum regime by considering the  properties in thermal equilibrium of a single anharmonic oscillator embedded in a single thermal reservoir. In case of $\kappa>0$ corresponding to a globally stable potential surface, this particularly allows to compare with numerically exact results from a SLED simulation with the full anharmonicity taken into account. Figure~\ref{fig:anham_comp_det_stoch} displays the variances $\llangle q^2\rrangle$ and $\llangle p^2\rrangle$ obtained from both the deterministic method presented here and from the SLED calculation which serves as a benchmark for the cumulant truncation. Apparently, the latter performs very accurately in the considered window of values for $\kappa$ for both phase space operators even for stronger dissipation. The variation of the position and momentum fluctuations caused by the increasing $\kappa$ is almost $50\%$ for weak coupling. Our approach covers this variation in accordance with the benchmark data. It is also seen (cf.\ the insets) that the perturbative treatment provides accurate results for higher order momements such as $\kappa\llangle q^4\rrangle$ and $\kappa\Bllangle\frac{qp^3+p^3q}{2}\Brrangle$, see (\ref{eq:moment_approx}). 

Note that in the considered range for $\kappa$ one also clearly sees the impact of friction on the variances: finite dissipation tends to suppress fluctuations in position and to enhance those in momentum. The cumulant formulation nicely captures this feature. 
The comparison with the exact data also allows us to quantify more precisely the accuracy of the perturbative treatment of the nonlinearities. For this purpose, according to (\ref{eq:Hamonean}), it is natural to consider the dimensionless quantity $\tilde{\kappa}= \kappa q_0^2/2 \omega^2$ with $q_0$ being a typical harmonic length scale as a measure for the relative impact of anharmonicities. In the low temperatures range of interest here, one chooses the ground state width of the harmonic system $q_0^2=\hbar/2m\omega$ which implies $\tilde{\kappa}=\hbar\kappa/4m\omega^2$. It then turns out that the perturbative treatment provides accurate results even  for $\tilde{\kappa}\sim 0.5$, i.e. not only for weak but also for moderately strong anharmonicities. 

%In terms of the effective frequency $\tilde{\omega}$, we find that the approach provides accurate results for $\tilde{\omega}^2-\omega^2=3\kappa \llangle q^2 \rrangle < \delta\times \omega^2$ with $\delta\approx 1.5$ and thus also for stronger anharmonicities. Note that in this domain the harmonic relations apply with effective frequencies though, i.e.,  $\llangle q^2\rrangle = (\hbar/2m\tilde{\omega}) {\rm coth}(\tilde{\omega}\hbar\beta/2)$ and $\llangle p^2\rrangle = (m\tilde{\omega}\hbar/2) {\rm coth}(\tilde{\omega}\hbar\beta/2)$.

In case of a softer mode ($\kappa<0$) a direct comparison with full numerical findings is no longer possible as the anharmonic potential is only locally stable. The perturbative treatment is thus physically sensible only as long as all relevant processes remain sufficiently localized around the minimum of the potential. In particular, the approach does not capture quantum tunneling through the potential barriers from the well into the continuum. However, it provides the nonlinearity required for rectification of heat transfer with the finite dissipation promoting localized states in the well. Figure~\ref{fig:anham_neg_kap} shows the phase space variances $\llangle q^2\rrangle$ and $\llangle p^2\rrangle$ together with the uncertainty product$\sqrt{\llangle q^2\rrangle\llangle p^2\rrangle}$. For very weak damping the latter remains at the minimal values 1/2, while position fluctuations increase and momentum fluctuations decrease with growing $|\kappa|$. This is in contrast to stronger friction, where already for $\kappa=0$ (harmonic limit) the uncertainty product exceeds 1/2 with strongly suppressed (enhanced) momentum (position) fluctuations for larger $|\kappa|$. Note that the range, where the approach is expected to be applicable, is restricted to substantially smaller values of $|\kappa|$ compared to the situation with $\kappa>0$. 
We close this analysis by presenting Wigner functions $W(q,p)$ for various values of $\kappa$ at $\kb T=0$, see figure.~\ref{fig:wigner_T_0}. The squeezing of phase space distributions is clearly seen with momentum squeezing for $\kappa<0$ and squeezing in position for $\kappa>0$. \newline

\bigskip
\textit{c) Towards quantum heat transfer: anharmonic chains}\newline

\begin{figure}
%\begin{minipage}{8.5cm}
%\includegraphics[width=7.5cm]{../figures/covarianzmatrix_10HO_kap_-0p10.pdf}
%\end{minipage}
%\vspace{20pt}
%\begin{minipage}{8.5cm}
%\includegraphics[width=7.5cm]{../figures/covarianzmatrix_10HO_kap_0p50.pdf}
%\end{minipage}
\begin{center}
\includegraphics[width=15cm]{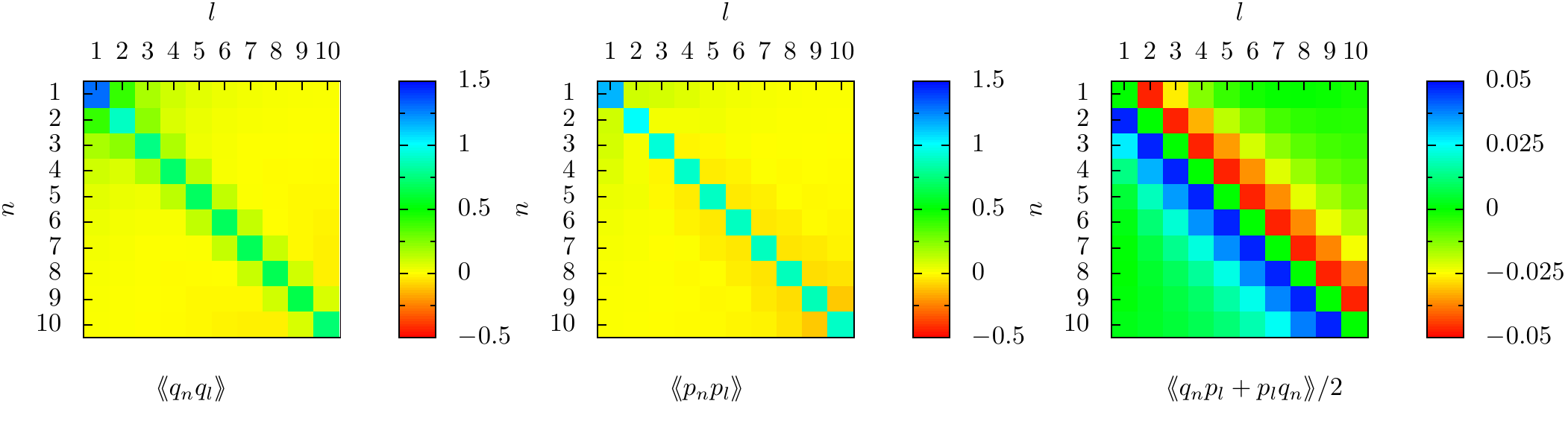}
\vfill
\vfill
\includegraphics[width=15cm]{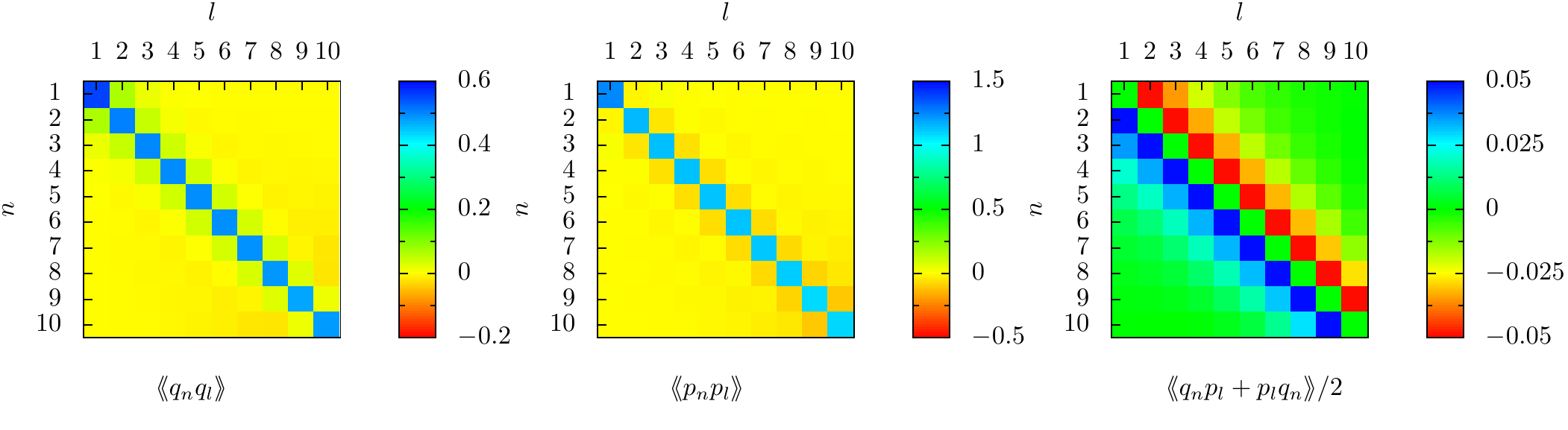}
\end{center}
\caption{Blocks of the covariance matrices $\BSig$ for ordered anharmonic chains with $\kappa=-0.10$ (top) and $\kappa=0.50$ (bottom). The columns show different parts of $\BSig$: all covariances of the positions $\llangle q_nq_l\rrangle$ (left), of the momenta $\llangle p_n p_l\rrangle$ (middle) and symmetrized mixed products $\llangle q_np_l+p_lq_n\rrangle/2$ (right). The chains accord to the Hamiltonian in (\ref{eq:Ham_qu}) with  $N=10$ oscillators. The first one $n=1$ is attached to a bath with $T_h=1$ and the last one $n=N=10$ to  a reservoir with  temperature $T_c=0.0$. Other parameters are $\gamma_l=\gamma_r=0.1$, $\mu=0.3$, $\omega=1$, $m=1$, $\hbar=1$ and cut-offs for both reservoirs are $\omega_c=100$.}
\label{fig:covmats_chains}
\end{figure}

One advantage of the developed methodology is its high computational efficiency also for large systems. Here, we analyze heat transfer through a chain of anharmonic oscillators of length $N=10$ according to (\ref{eq:Ham_qu})  terminated at both ends by thermal reservoirs at temperatures $T_l=T_h$ and $T_r=T_c<T_h$, respectively. We put all intra-oscillator couplings to $\mu=0.3$, masses to $m=1.0$ and frequencies to $\omega=1.0$; further $T_c=0$ while $T_h=1$ in dimensionless units. 

Density plots in figure~\ref{fig:covmats_chains} show the covariance matrices $\BSig$ for two anharmonicity parameters $\kappa=-0.10$ (top) and $\kappa=0.50$ (bottom). One clearly sees the impact of a temperature difference as the diagonal elements corresponding to the mode coupled to the hot bath are substantially larger than those attached to the cold one. The temperature dependence of the effective frequencies leads to a more distinct impact of the nonlinearity on the oscillators coupled to the hot bath. Apparently, for these sites the negative $\kappa$ leads to a broading of the state in position as compared to the case for positive $\kappa$.

\begin{figure*}
\begin{minipage}{8.5cm}
\includegraphics[width=7.5cm]{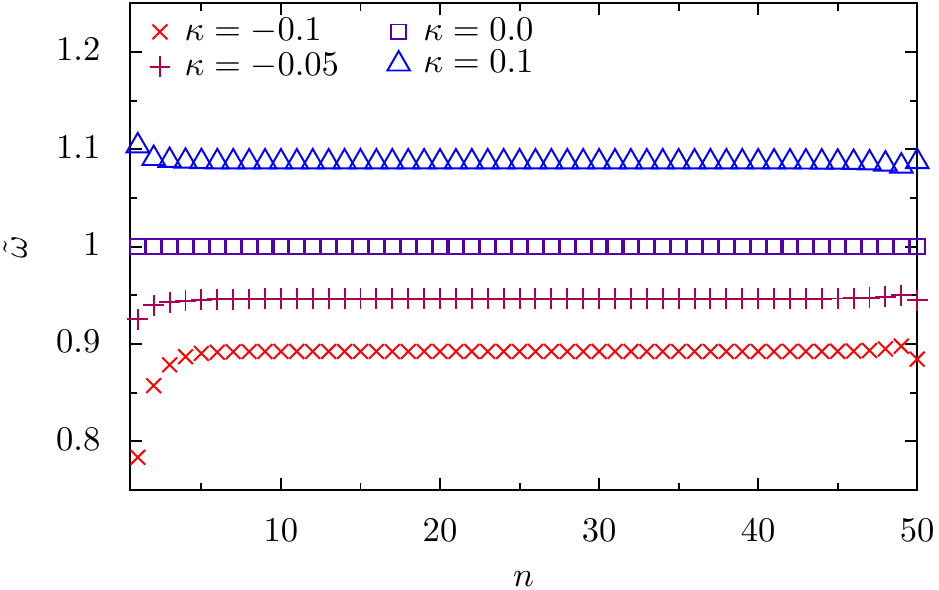}
\end{minipage}
\hfill
\begin{minipage}{8.5cm}
\includegraphics[width=7.5cm]{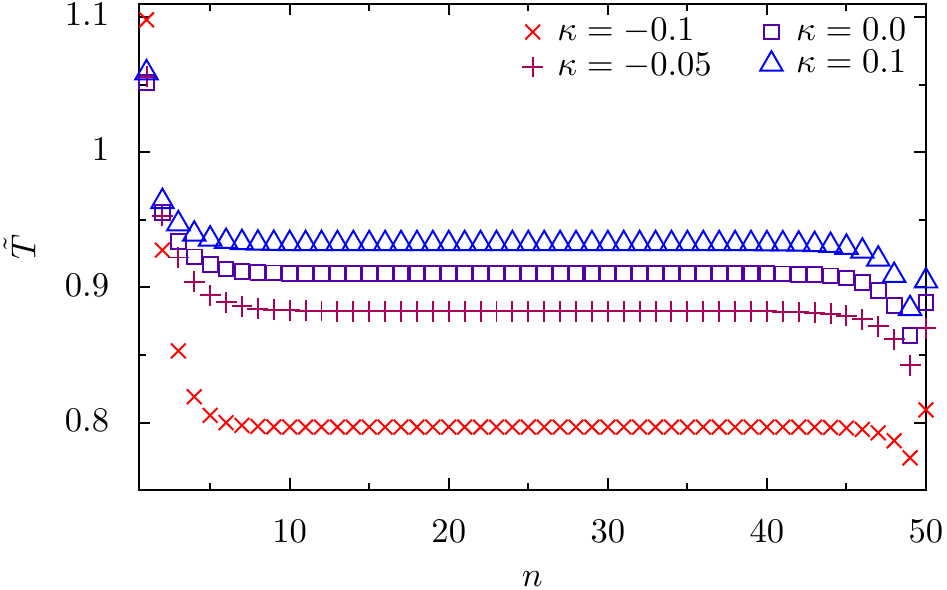}
\end{minipage}
\caption{The effective frequency $\tilde{\omega}$ (left) and temperature $\tilde{T}$ (right) over the site index $n$ of a chain with $N=50$ oscillators. Other model parameters are the same as in figure~\ref{fig:covmats_chains}.}
\label{fig:anham_effomtemp}
\end{figure*}

It is instructive to display the effective frequencies $\tilde{\omega}_n$ along the chain (cf.\ figure~\ref{fig:anham_effomtemp} left) for a long chain $N=50$. Apparently, the distribution is rather flat in the bulk and shows deviations only near the interfaces to the reservoirs with the frequencies being smaller (larger) for $\kappa<0$ ($\kappa>0$). We also show in figure~\ref{fig:anham_effomtemp} (right) the distribution of effective temperatures $\tilde{T}_n$ reconstructed from $\llangle p^2_n\rrangle=(m\tilde{\omega}_n\hbar/2)\coth[\hbar\tilde{\omega}_n/(2\kb \tilde{T}_n)]$. 
For further discussions of thermometry for dissipative systems we refer to very recent literature such as in ref.~\cite{Hovhannisyan2018}. 
Figure \ref{fig:anham_effomtemp} reveals a similar profile as that for the effective frequencies, thus indicating ballistic transport. This is confirmed by results for the heat currents (not shown), which do not reveal any dependence on the chain length. The fact that the temperature difference drops only within narrow interfaces between phonon chain and reservoir is reminiscent of the voltage drop in molecular chains in contact to electronic leads \cite{Cuevas}.

The question to what extent anharmonicities alone may lead to normal heat flow has led to conflicting results recently depending on the model under consideration \cite{Lepri1998, Terraneo2002}.  For the weak to moderate anharmonicities considered here, a combination of nonlinearity and disorder such as studied for example in \cite{Li_Prosen2004, Dhar2008} may provide a mechanism to induce normal heat flow. This is, however, beyond the scope of the current study, where we focus on heat rectification, and will be explored elsewhere.

\section{Rectification of quantum heat transport: single oscillator}
\label{sec:rect_1HO}

One of the simplest models to realize rectification consists of an anharmonic oscillator coupled to two reservoirs at different temperatures $T_l$ and $T_r$. The only way to induce a spacial asymmetry in such a model is to choose different coupling strengths $\gamma_l$ and $\gamma_r$ which remain constant under an exchange of reservoir temperatures. In steady-state, we are free to choose any of the two currents between system and reservoirs as they have identical absolute values. The energy current between the left reservoir and the oscillator reads according to (\ref{eq:heatfluxres1})
\[
\llangle j_{l,1}\rrangle = \frac{y_p^{(l)}}{m}- \gamma_l\Bllangle\frac{p^2}{m}\Brrangle
\]
with $y_p^{(l)}$ being the momentum part of the system-bath correlation function corresponding to the left reservoir. 

In the classical limit, this is $y_p^{(l)}=m\gamma_l/\beta_l$, while $\llangle p^2\rrangle$ for a system coupled to two baths is determined by additive fluctuations and damping strengths, respectively, i.e.,  
\[
\llangle p^2\rrangle=\frac{y_p^{(l)}+y_p^{(r)}}{\gamma_l+\gamma_r}\,.
\]
The resulting heat current reads 
\[
\llangle j_{l,1}\rrangle = \frac{\gamma_l\gamma_r}{\gamma_l+\gamma_r}\kb\Delta T
\]
with $\Delta T=T_r-T_l$. This immediately shows that {\em no} rectification can be observed for a single classical mode since an exchange of temperatures would result in a current with identical absolute value \mbox{$\jhc = -\jch$}.

\begin{figure}
\begin{center}
\includegraphics[width=8.0cm]{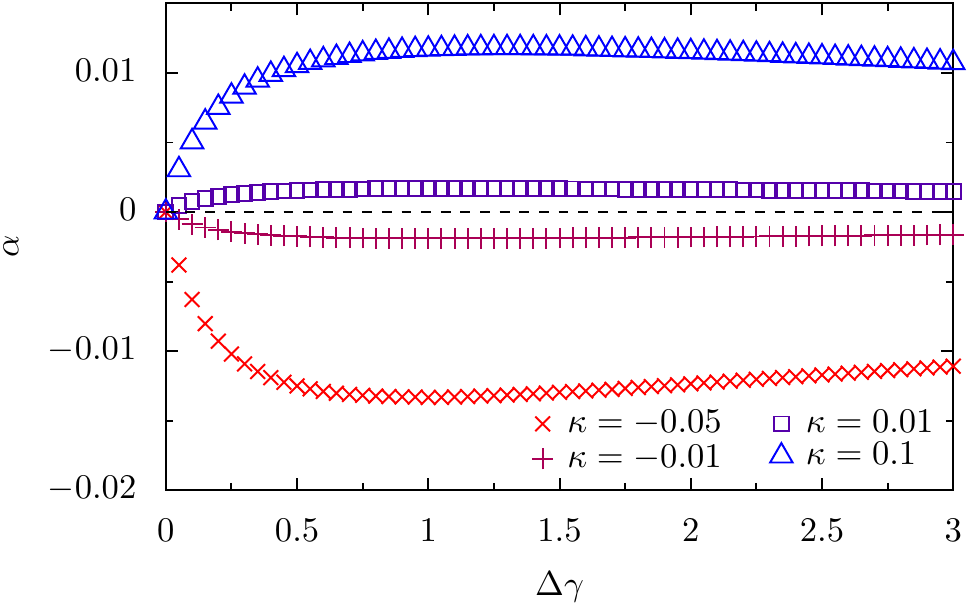}
\end{center}
\caption{Rectification coefficient $\alpha$ versus the difference of damping strengths $\Delta\gamma = \gamma_r-\gamma_l$ for a single anharmonic oscillator coupled to two quantum reservoirs with constant $\gamma_l=0.1$ and varying $\gamma_r$. The reservoirs have temperatures $T_h=1$  $T_c=0.0$ and cut-off frequencies $\omega_c=100$. Other parameters are $\omega=1$, $m=1$ and $\hbar=1$.}
\label{fig:rect_oneHO}
\end{figure}

The previous argument, however, does not apply to the quantum case, where the system-reservoir correlations depend on details of the Green's function caused by the non-Markovianity of the reservoirs. This Green's function is calculated with the effective frequency $\tilde{\omega}$ which varies with an exchange of the reservoir temperatures.

Therefore, rectification is possible with a single nonlinear system degree of freedom as shown in refs.~\cite{Segal2005, Ruokola2009}. Figure~\ref{fig:rect_oneHO} shows the rectification coefficient $\alpha$ versus $\Delta\gamma = \gamma_r-\gamma_l$ where $\gamma_l=0.1$ is kept constant. Results for positive and negative values of $\kappa$ over the whole range of $\Delta\gamma$ reveal that $\kappa <0$ leads to $\alpha<0$ and $\kappa>0$ to $\alpha >0$. This difference in the signs can be attributed to the different level structure of the system, where for the softer mode the energy level spacings are narrower for higher lying states while the opposite is true for the stiffer mode, see figure~\ref{fig:pot_anharm}.

\section{Rectification of  quantum heat transport: chains of oscillators}\label{sec:rect_chains}

\begin{figure*}
\begin{minipage}{8.5cm}
\includegraphics[width=7.5cm]{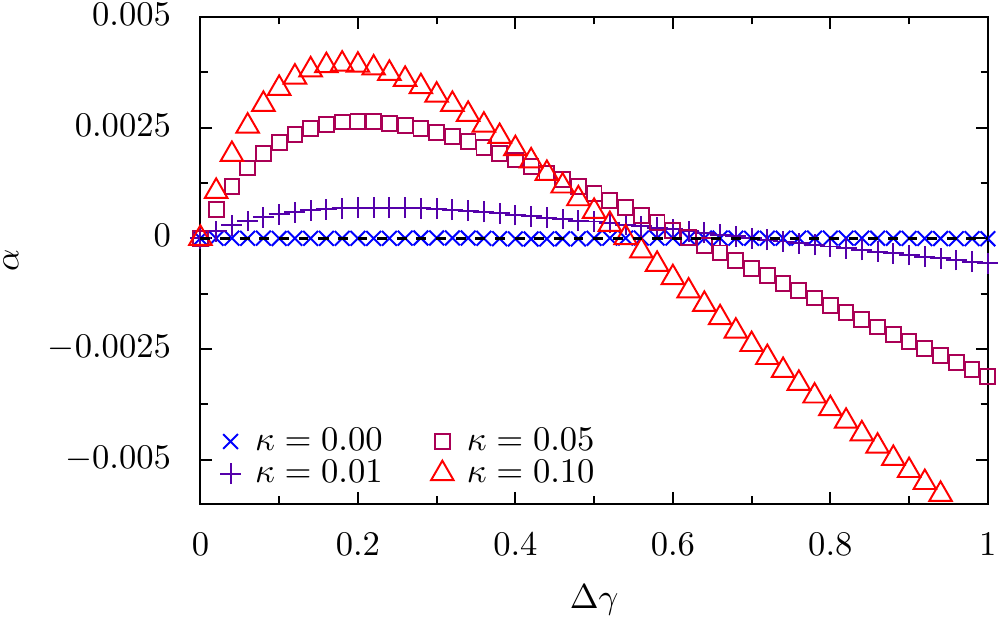}
\end{minipage}
\hfill
\begin{minipage}{8.5cm}
\includegraphics[width=7.5cm]{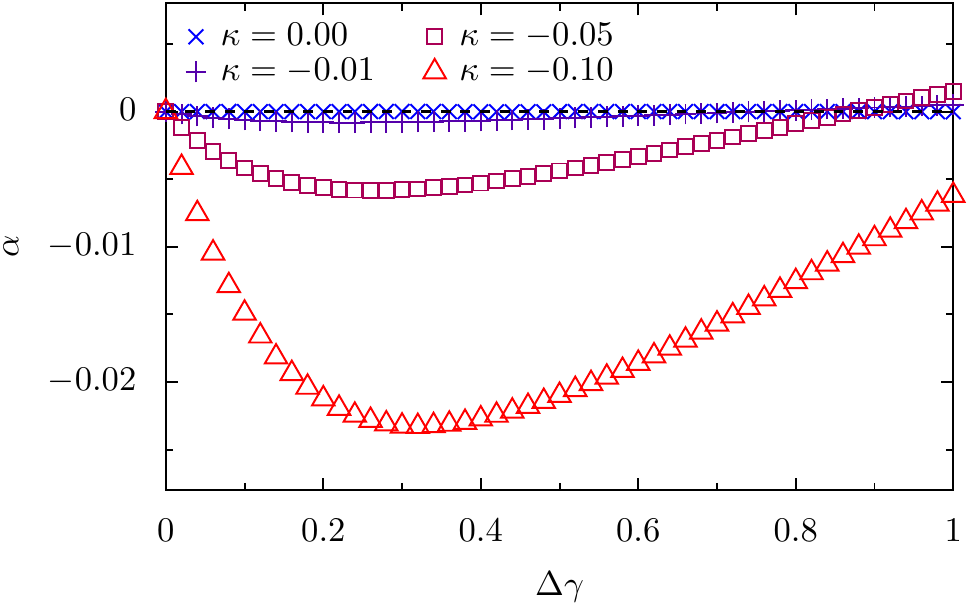}
\end{minipage}
\caption{Rectification coefficient $\alpha$ over $\Delta\gamma=\gamma_r-\gamma_l$ for positive $\kappa$ (left) and negative (right) for a system consisting of two anharmonic oscillators according to (\ref{eq:Ham_qu}) for $N=2$. $\gamma_l=0.1$ is constant, while $\gamma_r$ is varied. Each oscillator is coupled to its own reservoir which have different temperatures $T_h=1$, $T_c=0$ and equal $\omega_c=100$. Other parameters are: $\mu=0.3$, $\omega=1$, $m=1$, $\hbar=1$.}
\label{fig:rect_overgam_varkappa}
\end{figure*}

Going from a single oscillator to chains of oscillators is not just a quantitative modification of the setting. It is particularly interesting as it allows, by tuning the asymmetry either in the coupling to the reservoirs or in the on-site frequencies, to control the rectification of being positive or negative. This may be of relevance for experimental realizations of heat valves as they have very recently been explored in \cite{Ronzani2018}. Underlying physical principles can be revealed already for a set-up consisting of two oscillators, where we keep frequency and coupling strength of the oscillator at site $n=1$ fixed and vary those at $n=2$, i.e.\ $\omega_2=\omega_1+\Delta\omega$ and $\gamma_r=\gamma_l+\Delta\gamma$.  

Before we discuss specific results below, let us already here elucidate the main mechanism that governs the rectification process. For the Hilbert space of the oscillator chain alone (no reservoirs) two sets of basis functions are distinguished, namely, the one consisting of the localized eigenstates of the individual oscillators and the one consisting of the normal modes of the coupled system. Then, roughly speaking, if the inter-oscillator coupling $\mu$ dominates against the oscillator-reservoir couplings $\gamma_l,\gamma_r$,  heat transfer occurs along the delocalized normal modes, while in the opposite situation localized states rule the game. It turns out that the changeover from one regime to the other one by tuning either $\Delta\omega$ or $\Delta\gamma$ is associated with a sign change in the rectification coefficient $\alpha$. This also implies that starting from the symmetric situation a growing asymmetry first leads to an extremum  of $\alpha$ before one reaches the point $\alpha=0$. 

\begin{figure*}
\begin{minipage}{8.5cm}
\includegraphics[width=7.5cm]{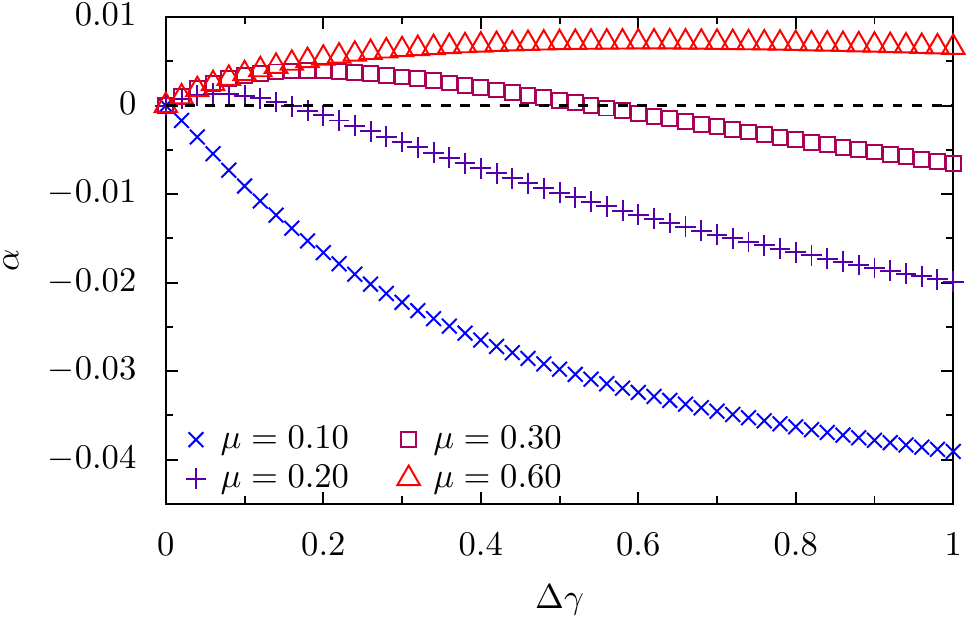}
\end{minipage}
\hfill
\begin{minipage}{8.5cm}
\includegraphics[width=7.5cm]{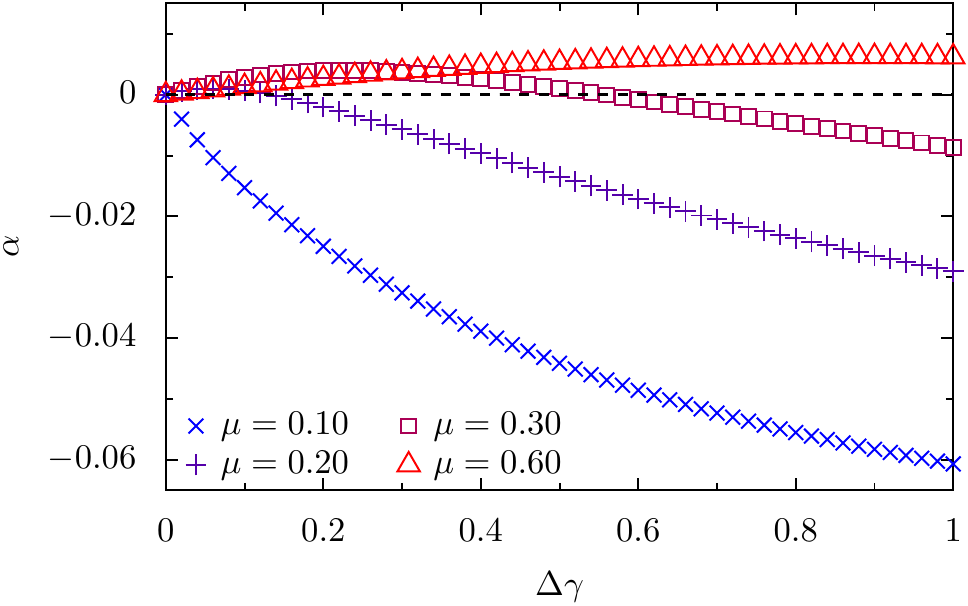}
\end{minipage}
\caption{Rectification coefficient $\alpha$ over $\Delta\gamma$ for positive $\kappa$  from a quantum (left) and a classical (right) model with delta correlated (Markovian) reservoirs. The models have similar parameters like that from figure~\ref{fig:rect_overgam_varkappa} (left) but with different intra-oscillator couplings $\mu$. The cut-off frequencies of the quantum reservoirs are $\omega_c=100$ and the temperatures for classical and quantum baths are $T_h=1$ and $T_c=0$.}
\label{fig:rect_overeta_varmu}
\end{figure*}

This discussion can be made more quantitatively by considering the effective spectral density $J_{\mathrm{eff},r}(\omega)$ of the right reservoir as seen from the first oscillator (fixed parameters) through the second one (varying parameters). For this purpose, we employ the formalism developed in \cite{Garg1985} and obtain for a purely ohmic bath [limit in (\ref{eq:spectralden}) for $\omega_c\to \infty$] the expression
\begin{equation}
J_{\mathrm{eff},r}(\omega)=\bar{\mu}^2\, \frac{m\omega\gamma_r \tilde{\omega}_2^4}{(\tilde{\omega}^2_2-\omega^2)^2+4\omega^2\gamma_r^2}\,,
\label{eq:eff_spect}
\end{equation}
with the dimensionless inter-oscillator coupling $\bar{\mu}=\mu/m\tilde{\omega}_2^2$ and effective frequencies $\tilde{\omega}_{1/2}^2=\omega_{1/2}^2+3\kappa\llangle q_{1/2}^2\rrangle$. It has the expected Lorentzian form with a maximum around $\omega=\tilde{\omega}_2$ and reduces
in the low frequency regime to an ohmic type of density $J_{\mathrm{eff},r}(\omega\to 0)\approx \bar{\mu}^2 \, m \omega \gamma_r$.  Note that the effective  spectral density (\ref{eq:eff_spect}) contains all three relevant parameters: the inter-oscillator coupling $\mu$ as well as   $\gamma_r$ and $\omega_2$, which we use to induce spatial asymmetry to the system. 

Now, let us consider the case with $\Delta\omega=0$ and varying $\Delta\gamma$ such that in the symmetric situation,  $\Delta\gamma=0$,  one has\ $\mu > \gamma_l, \gamma_r$. As a consequence, for growing $\Delta\gamma$ first the delocalized mode picture applies and the rectification coefficient approaches an extremum around that asymmetry point, where heat transfer at resonance  is optimal due to a matching of reservoir coupling constants to oscillator 1, i.e.\ $\gamma_{\mathrm{eff},r}\approx \gamma_l$ with the effective coupling from oscillator 1 to the right reservoir $\gamma_{\mathrm{eff},r}\equiv J_{\mathrm{eff},r}(\omega=\tilde{\omega}_1\approx \tilde{\omega}_2)/m\tilde{\omega}_2\equiv \bar{\mu}^2\, \tilde{\omega}_2^2/4\gamma_r$, i.e.,
\begin{equation}
\bar{\mu}^2 \tilde{\omega}_2^2 \approx 4\gamma_l \gamma_r\, .
\label{eq:rectmatching}
\end{equation}
With further increasing $\Delta\gamma$ the rectification tends to zero, changes sign, the asymmetry dominates against the inter-oscillator coupling, and a picture based on localized modes captures the heat transfer. We will see below that the sign of the rectification coefficient $\alpha$ also depends on the sign of the anharmonicity parameter $\kappa$.

\bigskip
\textit{a) Rectification by variation of the damping}\newline

Detailed results are now first discussed for the situation, where asymmetry is induced by a varying damping, see Figure~\ref{fig:rect_overgam_varkappa}. As anticipated, one indeed observes a change in the sign of the rectification coefficient together with an extremum for moderate asymmetries. When comparing the location of the extrema with the prediction according to the matching condition (\ref{eq:rectmatching}), one finds an excellent agreement. For the chosen parameters $\mu>\gamma_l, \gamma_r$ in the symmetric case, the previous discussion directly applies. The sign change in $\alpha$ can also be understood from the impact of friction onto $\llangle q_{1/2}^2\rrangle$ and thus onto the effective frequencies $\tilde{\omega}_{1/2}$: For strong coupling to the right reservoir ($\gamma_l\ll \gamma_r$) one always has (for identical harmonic frequencies) $\tilde{\omega}_2 < \tilde{\omega}_1$ for $\kappa>0$ and $\tilde{\omega}_2 > \tilde{\omega}_1$ for $\kappa<0$  due to the strong squeezing in position of oscillator 2. Quantum mechanically, the different energy level spacings then lead to $|\jch|>|\jhc|$ and thus $\alpha<0$ for $\kappa>0$ and $|\jch|<|\jhc|$ with $\alpha>0$ for $\kappa<0$. For weak asymmetry ($\gamma_l \leq\gamma_r$) the normal modes couple slightly more efficient to the right reservoir and the bottleneck is the coupling to the left reservoir. Consequently, the relation between the respective heat currents and the sign of $\alpha$ interchanges compared to the strongly asymmetric situation.  

 It is now also instructive to look for the rectification when the inter-oscillator coupling is tuned, see  figure~\ref{fig:rect_overeta_varmu}: Following our above reasoning the larger $\mu$, the stronger needs the asymmetry to be in order to induce a sign change in the rectification, while for very weak coupling no sign change occurs at all.  The mechanism developed above, namely, delocalized modes versus localized modes depending on the strength of the asymmetry, suggests that for purely classical reservoirs a qualitative similar behavior should be present. This is indeed the case as figure~\ref{fig:rect_overeta_varmu} (left) reveals. 
 
\begin{figure*}
\begin{minipage}{8.5cm}
\includegraphics[width=7.5cm]{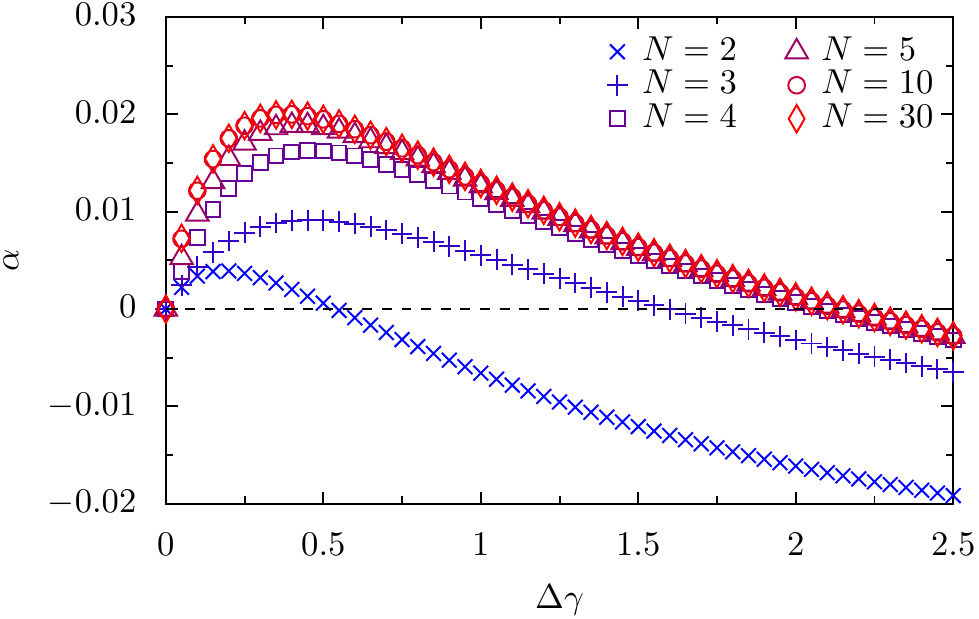}
\end{minipage}
\hfill
\begin{minipage}{8.5cm}
\includegraphics[width=7.5cm]{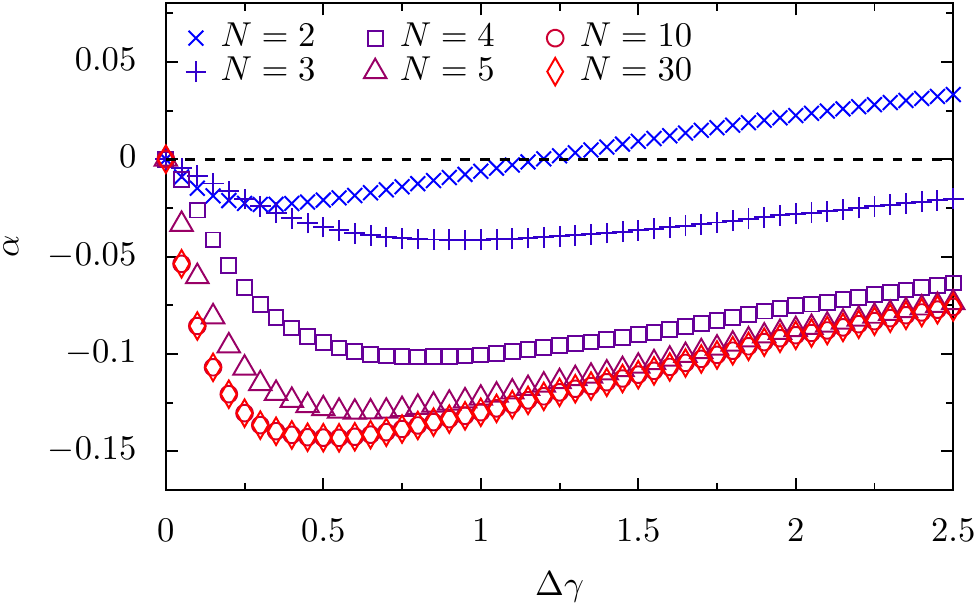}
\end{minipage}
\caption{Rectification coefficient $\alpha$ over $\Delta\gamma$ for positive $\kappa=0.10$ (left) and negative $\kappa=-0.10$ (right). The number of oscillators $N$ is varying, while all other parameters are as in figure~\ref{fig:rect_overgam_varkappa}.}
\label{fig:rect_overeta_varN}
\end{figure*}

 The impact of the number of oscillators $N$ in the chain on the rectification is displayed in figure~\ref{fig:rect_overeta_varN}. With growing $N$ the asymmetry must increase as well in order to induce a sign change in $\alpha$. We ascribe this to the feature already addressed above (cf.~figure~\ref{fig:anham_effomtemp}), namely, an effective screening of the asymmetry such that the bulk remains robust against variations of the coupling to the right reservoir. However, the absolute value of $\alpha$ increases with increasing chain length. In this and the previous cases rectification up 10\%-15\% can be seen in agreement with what has been found in the quantum regime in other studies \cite{Ruokola2009}. This also applies to situations where the frequency is modulated (next section) and reflects the fact that in the low energy sector (low temperatures) quantum oscillators are less influenced by anharmonicties than for higher lying states. 
 
\bigskip
\textit{b) Rectification by modulation of the frequencies}\newline

\begin{figure}
\begin{minipage}{8.5cm}
\includegraphics[width=7.5cm]{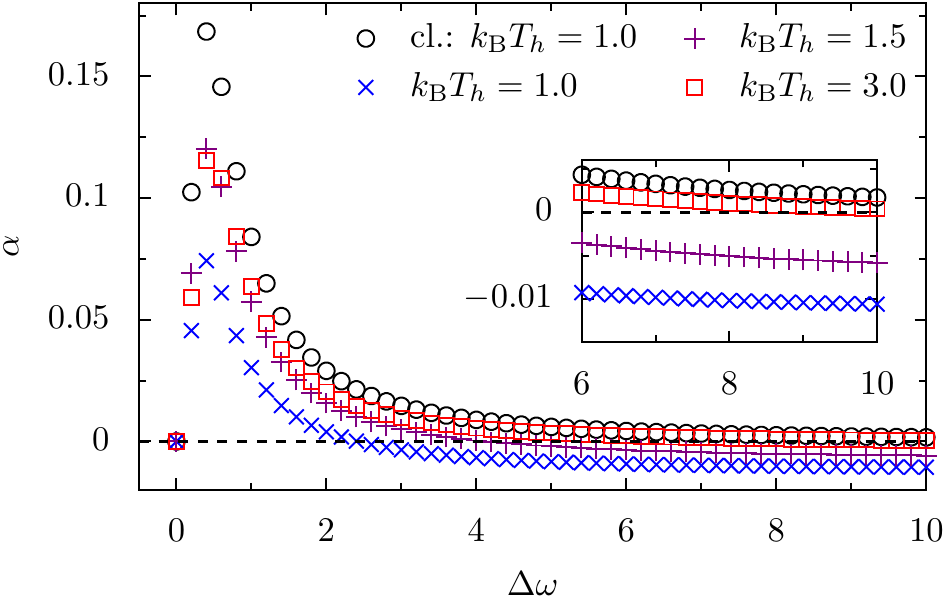}
\end{minipage}
\hfill
\begin{minipage}{8.5cm}
\includegraphics[width=7.5cm]{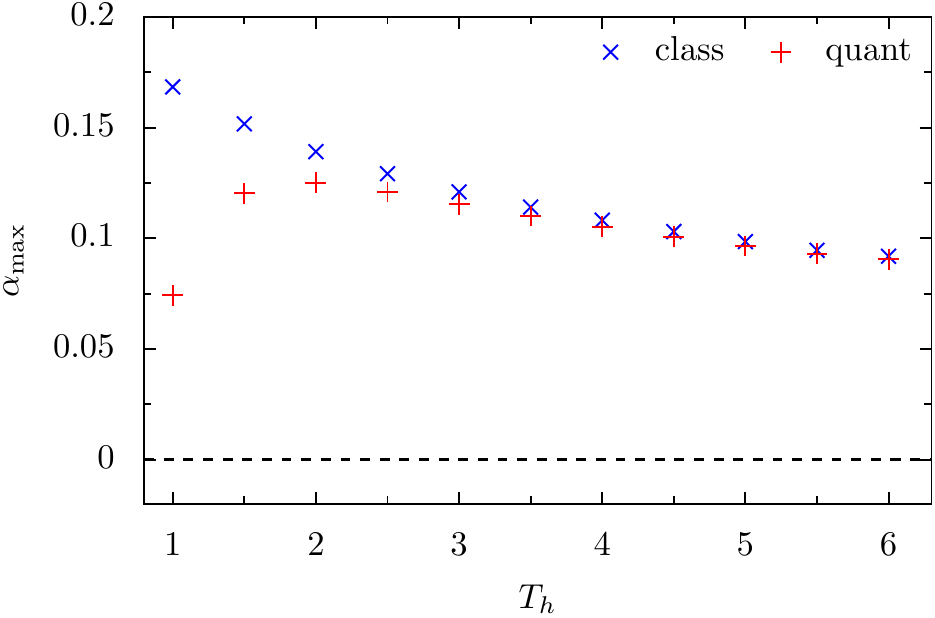}
\end{minipage}
\caption{Left: Rectification coefficient $\alpha$ versus the difference of the frequencies $\Delta\omega = \omega_2-\omega_1$ with $\omega_1=1.0$ for a system of two coupled modes according to (\ref{eq:Ham_qu}) with $N=2$. The system is terminated by  reservoirs with $\Delta T= T_h-T_c=1.0$, while different combinations of $T_h$ and $T_c$ are shown. The black circles show $\alpha$ for classical delta correlated reservoirs. Right: The rectification versus the temperature of the hot bath $T_h$ for $\Delta T=1.0$ and $\omega_2=1.4$ which is the value where $\alpha$ is maximal. Both plots: The cut-off frequency of the quantum reservoirs is $\omega_c=100$ and the dampings are constant $\gamma_l=\gamma_r=0.1$. The intra-oscillator coupling is $\mu=0.3$ and both modes are anharmonic with $\kappa=0.1$. $\alpha <0$ is a quantum feature which occurs at low temperatures. Other parameters are $m=1$ and $\hbar=1$.}
\label{fig:rect_varomega_diffT}
\end{figure}

\begin{figure*}
\begin{minipage}{8.5cm}
\includegraphics[width=7.0cm]{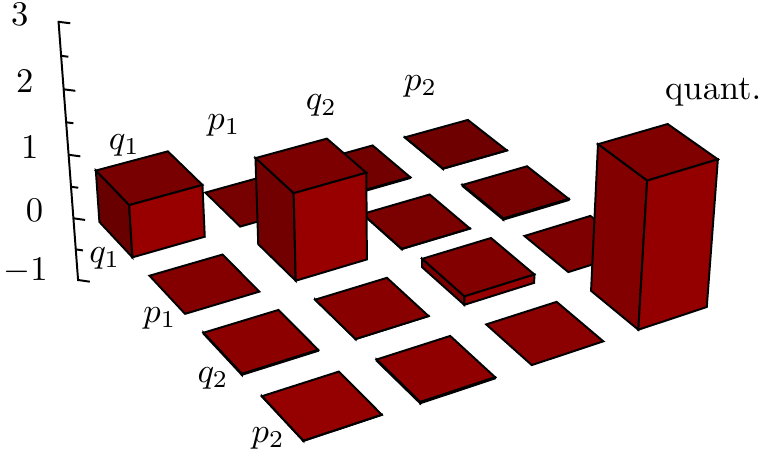}
\end{minipage}
\hfill
\begin{minipage}{8.5cm}
\includegraphics[width=7.0cm]{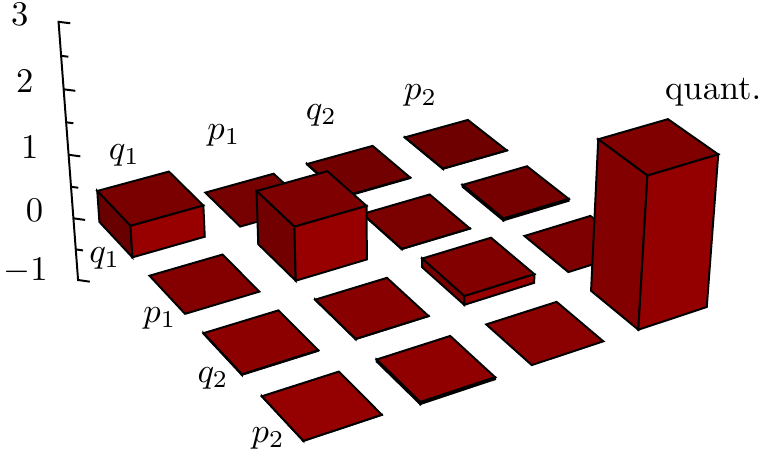}
\end{minipage}
\vfill
\begin{minipage}{8.5cm}
\includegraphics[width=7.0cm]{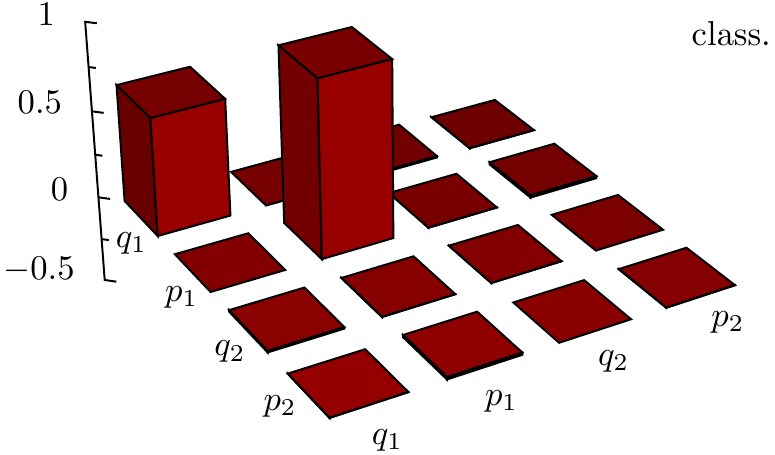}
\end{minipage}
\hfill
\begin{minipage}{8.5cm}
\includegraphics[width=7.0cm]{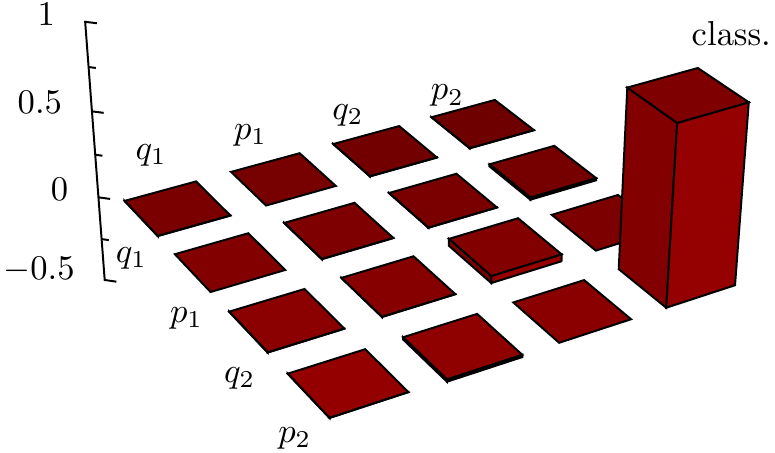}
\end{minipage}
\caption{The steady-state covariance matrix $\BSig$ for the system from figure~\ref{fig:rect_varomega_diffT} for both configurations $h\rightarrow c$ (left) and $c\leftarrow h$ (right) with quantum (top) and classical baths (bottom). The matrices are plotted for $\Delta\omega=4.0$ and reservoir temperatures $T_h=1$ and $T_c=0.0$. These parameters lead to $\alpha <0$ for the quantum case which can be explained by a tunneling between the modes for $c\leftarrow h$ where both modes are in the ground state (right). Instead, the first mode for $h\rightarrow c$ experiences thermal fluctuations (left) which reduces the overlap of both wave functions.}\label{fig:covmat_bar3d}
\end{figure*}

As an alternative to a variation of the chain-reservoir coupling, spatial asymmetry can also be induced by varying on-site frequencies $\omega_n$. Figure~\ref{fig:rect_varomega_diffT} shows $\alpha$ for a system of two coupled anharmonic oscillators with the asymmetry being quantified by $\Delta\omega=\omega_2-\omega_1$  with $\omega_1=1.0$ put constant, while $\omega_2$ is tuned. Instead of analyzing  different values for the anharmonicity parameter as above, we here consider a constant temperature difference at different individual temperatures from  the quantum up to the classical regime. One again observes a distinct maximum for $\alpha$, whose value $\alpha_\mathrm{max}$ is depicted in figure~\ref{fig:rect_varomega_diffT} (right) when $T_h$ is varied. The positive $\alpha$ for moderate $\Delta\omega$ can be understood by noting that $\llangle q_n\rrangle$ is larger if the $n$-th mode is coupled to the hot bath which in turn increases the effective frequency $\tilde{\omega}_n$ if $\kappa>0$. For our setting this means that the coupling of the hot bath for $h\rightarrow c$ to the first mode compensates the frequency increase of the second mode for finite $\Delta\omega$. For $c\leftarrow h$ instead, the coupling of the hot bath to the second mode amplifies the detuning caused by $\Delta\omega$. For the quantum case, this effect is less distinct as ground state fluctuations act also on the mode which is coupled to the cold bath (see figure~\ref{fig:covmat_bar3d}). As already mentioned in the previous section, typical values of the rectification for the models considered here are on the order of 10\% and do not exceed 15\%-20\% in accordance with results reported in \cite{Segal2005}. In fact, even for the extreme case of a single two level system, rectification was found to be on the order of 10\% \cite{Ruokola2009}. It would be interesting to explore whether this can be improved by particular designs of chains. We emphasize that figure~\ref{fig:rect_varomega_diffT} (right) reveals that our approach also provides finite rectification in the high temperature range, where it matches the corresponding classical predictions, a non-trivial test for the consistency of cumulant-type of expansions (see \cite{Bricmont2007}).

What is more striking though is the behaviour for larger asymmetries: While the classical rectification approaches zero from above, in the low temperature quantum regime we see again a change in sign of $\alpha$ and then a saturation with further increasing $\Delta\omega$.  This is a genuine quantum phenomenon as an inspection of the covariance matrix reveals, see  figure~\ref{fig:covmat_bar3d}. Classically, the phase space entries for the oscillator coupled to the reservoir at $T_c=0$ are absent, while this is not the case quantum mechanically. In the latter case, ground state fluctuations survive, where those of the oscillator with larger frequency $\tilde{\omega}_2$ exceed those of the softer one. This then gives rise to the observed finite rectification also for large $\Delta\omega$. Why does the rectification saturate in this limit? To understand this we again exploit the effective spectral density specified in (\ref{eq:eff_spect}): For $\omega_2\gg \omega_1$ due to $\tilde{\omega}_2\gg \tilde{\omega}_1$ the first oscillator effectively probes only the ohmic-type low frequency portion of the distribution $J_{\mathrm{eff},r}(\omega\ll \tilde{\omega}_2)\approx \bar{\mu}^2 \, m \omega \gamma_r$ which is independent of $\omega_2$. Heat transfer is thus governed by low frequency quantum fluctuations, a process that may also be interpreted as quantum tunneling through the second oscillator.

\begin{figure}
\begin{center}
\includegraphics[width=8.0cm]{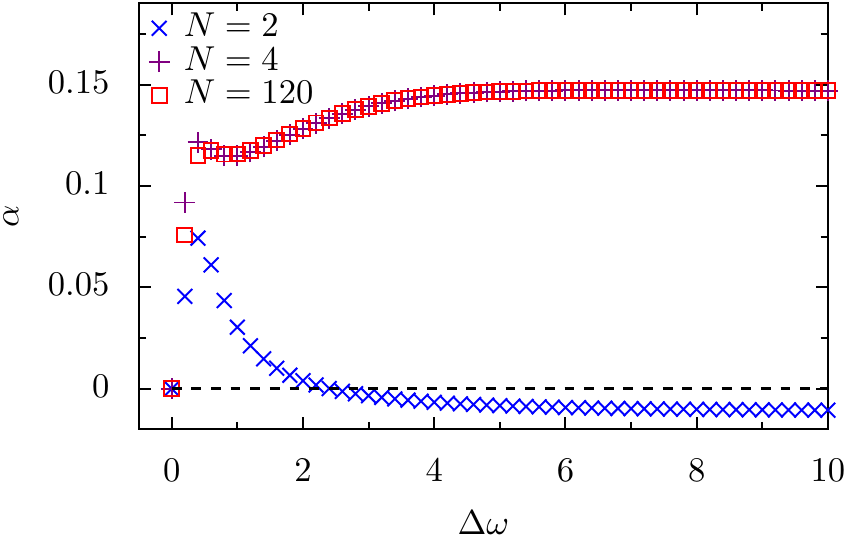}
\end{center}
\caption{Rectification $\alpha$ for a series connection of heat valves terminated by quantum reservoirs. The oscillators with odd index have low frequencies $\omega_\mathrm{low}=1.0$, the oscillators with even index have high frequencies $\omega_\mathrm{high}=\Delta\omega+\omega_\mathrm{low}$. Other parameters are $\kappa=0.1$, $\mu=0.3$, $m=1$, $\hbar=1$, $T_h=1$, $T_c=0.0$, $\gamma_l=\gamma_r=0.1$ and $\omega_c=100$ for both reservoirs.
}
\label{fig:rect_overomega_series}
\end{figure}
The detuning of the oscillator's frequency as a resource for rectification suggests an extension to a chain consisting of an even number of oscillators where the frequencies with even index are set to $\omega_\mathrm{high}$ and are varied, while the ones with odd index are kept constant at $\omega_\mathrm{low}=1.0$. Such a setting can be considered as a chain of heat valves, where one element consists of two coupled oscillators with $\omega_\mathrm{low}$ and $\omega_\mathrm{high}$, respectively. 
Figure~\ref{fig:rect_overomega_series} shows the rectification for such a series connection of heat valves. Apparently, $\alpha < 0$ is only obtained for one single valve with $N=2$, while a series of multiple valves leads to $\alpha > 0$, also for large $\Delta\omega$ and independently of the number of oscillators $N$. To understand the reason for this quite different behavoir, we consider the situation with two valve elements ($N=4$): Then, we have for $h\rightarrow c$ a significant increase of $\tilde{\omega}_1$ (strong fluctuations in position) which supports heat transport through the whole chain since then also  $\tilde{\omega}_2$ is affected by the hot bath. Instead, for $c\leftarrow h$ even a small $\Delta\omega$ is sufficient to screen the effect of the hot bath on $\tilde{\omega}_3$ so that this respective oscillator remains in the ground state. Therefore, the detuning between the oscillators $n=4$ and $n=3$ is larger than for $h\rightarrow c$ which implies $\alpha >0$ for chains with $N\geq 4$. In other words, the length dependence for $N\geq 4$ is determined by the presence of detuned modes in the bulk which are not attached to a reservoir, while the total number of those modes is not important.\newline

\section{Rectification in disordered systems}
\label{sec:rect_disordered}

\begin{figure*}
\begin{minipage}{8.5cm}
\includegraphics[width=7.5cm]{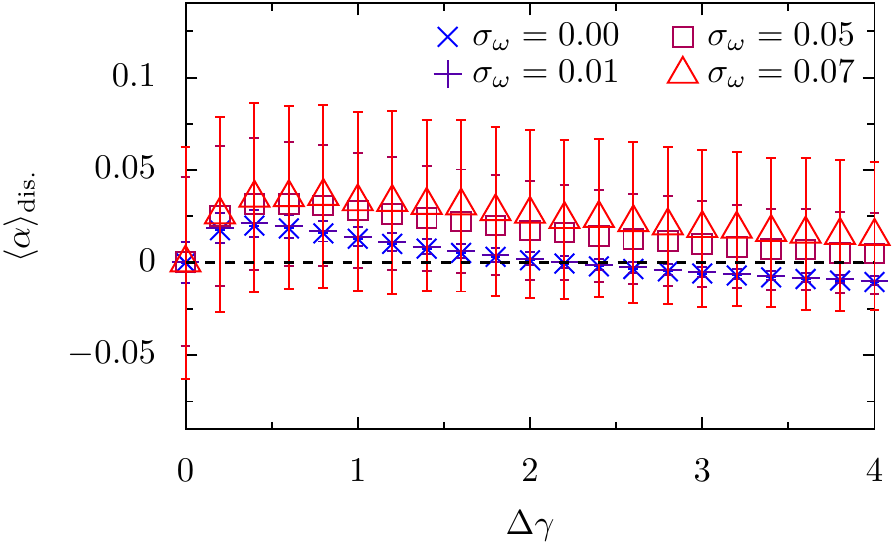}
\end{minipage}
\hfill
\begin{minipage}{8.5cm}
\includegraphics[width=7.5cm]{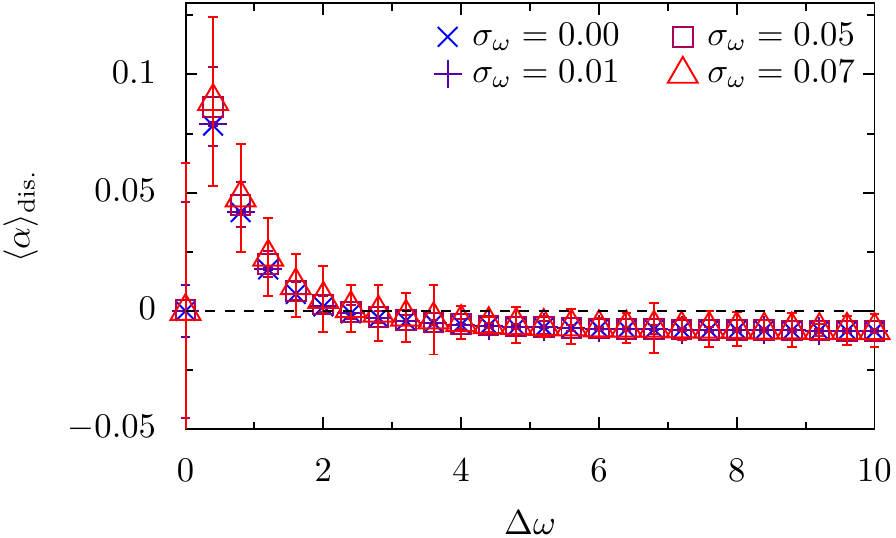}
\end{minipage}
\caption{Averaged rectification $\langle\alpha\rangledis$ for $3\times 10^3$ samples of random chains given by (\ref{eq:Ham_qu}) for $N=10$. The disorder is induced by normally distributed on-site frequencies with expectation values $\langle\omega_n\rangledis =1.0$, while the different colors show various standard deviations $\sigma_\omega$. Negative random values of $\omega_n$ are rejected. Other parameters are $\mu=0.3$, $m=1$, $\hbar=1$ and $T_h=1$, $T_c=0.0$, $\omega_c=100$ for both reservoirs, $\kappa=0.1$. The left panel shows a variation of the damping with $\Delta\gamma=\gamma_r-\gamma_l$ where $\gamma_l=0.1$ , while the on-site frequencies are homogeneous with $\langle\omega\rangledis=1.0$. The right panel shows varying frequencies $\langle\omega_N\rangledis$ with $\Delta\omega=\langle\omega_N\rangledis-\langle\opind{\omega}{low}\rangledis$ and $\langle\opind{\omega}{low}\rangledis=1.0$ representing all on-site frequencies but the last, while the dampings are equal $\gamma_l=\gamma_r=0.1$. The error bars represent the standard deviation of the distribution of the random $\alpha$.}
\label{fig:random_rect}
\end{figure*}

The rather delicate interplay of nonlinearity and spatial asymmetry which the rectification is based on rises the question of the stability of these effects in presence of disorder. In the sequel, we restrict ourselves to a randomization of the on-site frequencies $\omega_n$ and obtain the rectification as an average over several thousand of realizations, where the heat currents $\jhc$ and $\jch$ are calculated with identical realizations. In the left panel in figure~\ref{fig:random_rect} we present $\langle\alpha\rangledis$ for varying $\Delta\gamma=\gamma_r-\gamma_l$ and use normally distributed on-site frequencies with standard deviation $\sigma_\omega$ and mean $\langle \omega_n\rangledis=1.0$, where we reject negative values of $\omega_n$. We emphasize that $ \langle \omega_n\rangledis$ is an average with respect to random chains and \emph{not} over the individual sites $n$ of one chain. The right panel shows $\langle\alpha\rangledis$ for varying $\Delta\omega=\langle\omega_N\rangledis-\langle\opind{\omega}{low}\rangledis$ with normally distributed on-site frequencies and equal dampings $\gamma_l=\gamma_r=0.1$ on both ends.

Obviously, for varying $\Delta\gamma$ (left), the disorder frustrates the zero-crossing of the rectification observed at $\Delta\gamma\sim 2.0$ for the ordered case shown in figure~\ref{fig:rect_overeta_varN}. The profile of $\langle\alpha\rangledis$ for stronger disorder even suggests a convergence of $\langle\alpha\rangledis\to 0$ for large $\Delta\gamma$. For a system of two oscillators, we found that $\alpha<0$ arises from a slightly stronger detuning of the two frequencies for $h\rightarrow c$ than for $c\leftarrow h$. For the disordered case, this small effect seems to be washed out. Concerning the case of a varying $\Delta\omega$ shown in the right panel, it is interesting to observe that here disorder is less effective:
one has $\alpha <0$ for large $\Delta \omega$ also for stronger disorder.  Since in this regime $\Delta\omega$ is very large, it is clear that the system is more robust against  small variations of the $\omega_n$. This manifests as well  in the relatively small error bars for the random $\alpha$. Random couplings $\mu_n$ were also investigated and found to cause a similar behavior as random frequencies (not shown).

\bigskip

\section{Summary and Outlook}
\label{sec:sum_out}

In this paper we developed a framework to describe heat transfer in open quantum systems which also applies to strong thermal contact, allows to cover the full temperature range, various realizations (weak -- strong coupling, high -- low temperatures), and can be generalized to higher dimensions and other geometries. It thus may serve as platform to quantify the performance of heat valves, heat engines, or cooling devices as they are currently under study in atomic and mesoscopic physics. 

More specifically, the heat transfer and rectfication across chains of anharmonic oscillators has been explored, where the nonlinearity can be tuned from weak to moderately strong. Tuning the level of symmetry breaking by either changing the asymmetry in the chain-reservoir coupling or in the frequency distribution of the oscillators we find the rectification coefficient to pass from positive to negative or vice versa by running through an extremum. A deeper analysis reveals a mechanism, where heat is predominantly carried by non-local modes (weak symmetry breaking) or localized modes (strong symmetry breaking) with a smooth turnover between the two scenarios. Similar findings have been reported recently in an experimental realization of a heat valve based on superconducting circuits \cite{Ronzani2018}. While for symmetry breaking induced by the chain-reservoir coupling this mechanism also applies to the classical regime, a genuine quantum effect is found for symmetry breaking induced by frequency mismatch between adjacent oscillators. In this situation, strong symmetry breaking gives rise to a finite rectification (and thus a finite heat transfer) in contrast to the classical prediction. This finding may have relevance for recent proposals for the fast initializations of quantum bits (cooling) by frequency tuning \cite{Tuorila2017}.

\ack
Valuable discussion with J. Pekola, M. M\"ott\"onen, and R. Kosloff are gratefully acknowledged. Financial support was provided by the German Science Foundation through grant AN336/11-1, the Land Baden-W\"urttemberg through the LGFG program (M.W.),  and the Center for Integrated Quantum Science and Technology (IQST).

\appendix
\section{Steady-state equations for 2nd order trace cumulants}
\label{app:2ndtrace_cumulants}

By replacing $A_\xi$ with the product of $q$ and $p$ in the adjoint equation of the SLED $\difft A_\xi=\mathcal{L}^\dagger A_\xi$, equations of motion for the second order trace cumulants $\langle AB\ctrangle=\langle AB\trangle-\langle A\trangle\langle B\trangle$ can be calculated under usage of the product rule to account for the product of trace averaged operators. With the superoperator from (\ref{eq:adjoint_sled}) the following equations for the second cumulants are derived:

\begin{eqnarray}
&\frac{\rmd}{\rmd t}\langle q^2\ctrangle =\frac{2}{m}\Big\langle\frac{qp+pq}{2}\Big\ctrangle\nonumber\\
&\difft\langle p^2\ctrangle =-2m\omega^2\Big\langle\frac{qp+pq}{2}\Big\ctrangle-2\gamma\langle p^2\ctrangle\nonumber\\
&- 2m\kappa\Big[\Big\langle\frac{q^3p+pq^3}{2}\Big\trangle-\langle p\trangle\langle q^3\trangle\Big]\nonumber\\
&\difft\Big\langle\frac{qp+pq}{2}\Big\ctrangle =-m\omega^2\langle q^2\ctrangle-m\kappa[\langle q^4\trangle -\langle q^3\trangle\langle q\trangle]\nonumber\\
&+\frac{1}{m}\langle p^2\ctrangle-\gamma\Big\langle\frac{qp+pq}{2}\Big\ctrangle\,.
\label{eq:sec_moments_ex}
\end{eqnarray}
For the second and third equation, a calculation of commutators with powers of $q$ and $p$ according to $[q^k,p^l]=\rmi\hbar2k\{q^{k-1},p^{l-1}\}+pq^k p^{l-1}-p^{l-1}q^kp$ for $k=4,\,l=2$ and $[q^k,qp] = [q^k,pq]=k\rmi\hbar q^k$ for $k=4$ is used.\newline

\textit{a) Truncation of trace cumulants}\newline

The anharmonicity contributes via moments of higher order whose exact dynamics is not available. These moments expressed by cumulants read\newline
\begin{eqnarray}
&\langle q^3\trangle =\,\langle q^3\ctrangle + 3\langle q^2\ctrangle\langle q\ctrangle +\langle q\ctrangle^3\nonumber\\
&\Big\langle\frac{q^3p+pq^3}{2}\Big\trangle =\,\Big\langle\frac{q^3p+pq^3}{2}\Big\ctrangle +\langle q^3\ctrangle\langle p\ctrangle +3\langle q^2\ctrangle\Big\langle\frac{qp+pq}{2}\Big\ctrangle\nonumber\\
&+ 3\langle q\ctrangle\Big\langle\frac{q^2p+pq^2}{2}\Big\ctrangle+ 3\langle q^2\ctrangle\langle q\ctrangle\langle p\ctrangle\nonumber\\
&+ 3\langle q\ctrangle^2\Big\langle\frac{qp+pq}{2}\Big\ctrangle+\langle q\ctrangle^3\langle p\ctrangle\nonumber\\
&\langle q^4\trangle =\,\langle q^4\ctrangle+4\langle q^3\ctrangle\langle q\ctrangle+3\langle q^2\ctrangle^2\nonumber\\&+ 6\langle q^2\ctrangle\langle q\ctrangle^2+\langle q\ctrangle^4\,.
\label{eq:trace_moments_ex}
\end{eqnarray}
Setting all trace cumulants of order larger than two equal to zero and respecting the cancellation of some terms in the squared brackets in (\ref{eq:sec_moments_ex}) we arrive at

\begin{eqnarray}
&\difft\langle q^2\ctrangle =\frac{2}{m}\Big\langle\frac{qp+pq}{2}\Big\ctrangle\nonumber\\
&\difft\langle p^2\ctrangle =-2m\omega^2\Big\langle\frac{qp+pq}{2}\Big\ctrangle\nonumber\\
&-6m\kappa\Big[\langle q^2\ctrangle\Big\langle\frac{qp+pq}{2}\Big\ctrangle+\langle q\ctrangle^2\Big\langle\frac{qp+pq}{2}\Big\ctrangle\Big]-2\gamma\langle p^2\ctrangle\nonumber\\
&\difft\Big\langle\frac{qp+pq}{2}\Big\ctrangle =-m\omega^2\langle q^2\ctrangle-3m\kappa[\langle q^2\ctrangle^2+\langle q^2\ctrangle\langle q\ctrangle^2]\nonumber\\
&+\frac{1}{m}\langle p^2\ctrangle-\gamma\Big\langle\frac{qp+pq}{2}\Big\ctrangle\,.
\label{eq:sec_moments_tr}
\end{eqnarray}
this contains  only trace cumulants of first and second order what was to be achieved.\newline

It can be shown that this system of equations has a stable solution at zero for all trace cumulants, although $\langle q\ctrangle^2$ is a fluctuating quantity. To do so, we linearize (\ref{eq:sec_moments_tr}) by setting all products of second trace cumulants equal to zero and summarize the equations in a matrix-valued equation $\difft\vec{\sigma}_2=\mathbf{M}_2\vec{\sigma}_2$ with $\vec{\sigma}_2=(\langle q^2\ctrangle, \langle p^2\ctrangle, \Big\langle\frac{qp+pq}{2}\Big\ctrangle)^t$ and treat $\langle q\ctrangle^2$ as a parameter:
\begin{equation}
\mathbf{M}_2 =
\left(\begin{array}{ccc}
0 & 0 &2/m\\
0 & -2\gamma & -2m\omega^2-6m\kappa\langle q\ctrangle^2 \\
-m\omega^2-3m\kappa\langle q\ctrangle^2 & 1/m  & -\gamma 
\end{array}\right)\,.
\label{eq:M2tr}
\end{equation}
For positive $\kappa$, all eigenvalues have negative real part (stable fixed point). For negative $\kappa$ and weak anharmonicity, this condition is violated only with exponentially small probability. This may result in an extremely small drift, which will be neglected in a similar manner as tunneling out of the metastable well defined by negative $\kappa$. With this analysis, we managed to decouple the dynamics of the first- and second trace cumulants and can describe the steady-state covariance solely in terms of $\BSig^{\mathrm{(msc)}}$.

\section{Coefficients for dynamics of first trace cumulants in chain}
\label{app:system_matrix}

The matrix accounting for the dissipative dynamics of the system coordinates $q_n$ and $p_n$ determines the nonlinear motion via $a_n = m\tilde{\omega}_n^2 +2\mu$ for $2\leq n\leq N-1$ and $a_n = m\tilde{\omega}_n^2 +\mu$ for $n=1,N$:

\begin{equation}
\mathbf{M} =
\left(\begin{array}{cccccc}
0 & 1/m & 0 &\dots  & \dots & 0\\
-a_1 & -\gamma_l & \mu  & \dots & \dots & 0\\
\vdots  & \ddots & \ddots  &\ddots   &   & \vdots  \\
\vdots  &  &\ddots  & \ddots  &\ddots   & \vdots  \\
0 &\dots  & 0 & 0 & 0 & 1/m\\
0 &\dots  & \mu & 0 & -a_N & -\gamma_r
\end{array}\right)\,,
\label{eq:M1}
\end{equation}
with 
\[
\tilde{\omega}_n^2=\omega_n^2+3\kappa\llangle q_n^2\rrangle\,,
\]
being the effective frequency where we used $\llangle q_n\ctrangle\langle q_n\ctrangle\xrangle=\llangle q_n^2\rrangle$ which holds in the steady-state.

\section*{References}
\bibliography{literature_paper_rect}

\end{document}